\documentclass[%
aps, prl,
 amsmath,amssymb,
reprint,%
]{revtex4-1}

\usepackage{isomath}
\usepackage{amsmath,amsthm}
\usepackage{amsbsy}
\usepackage{amssymb}
\usepackage{amscd}
\usepackage{amsfonts}
\usepackage{stmaryrd}
\usepackage{siunitx}
\usepackage{euscript}
\usepackage[utf8]{inputenc}
\usepackage[T1]{fontenc}
\usepackage{newtxtext} 
\everymath{\displaystyle}
\usepackage{exscale}

\usepackage{graphicx}
\usepackage{boxedminipage}
\usepackage{calc}
\usepackage[usenames,dvipsnames]{xcolor}
\graphicspath{ {media/} }
\usepackage{setspace}
\usepackage{enumitem}
\setitemize{noitemsep,topsep=0pt,parsep=0pt,partopsep=0pt}
\setenumerate{noitemsep,topsep=0pt,parsep=0pt,partopsep=0pt}
\setdescription{noitemsep,topsep=0pt,parsep=0pt,partopsep=0pt}

\usepackage[colorinlistoftodos, color=green!40,prependcaption]{todonotes}

\usepackage{soul} % highlighting

\usepackage{xr-hyper}
\makeatletter
\newcommand*{\addFileDependency}[1]{
  \typeout{(#1)}
  \@addtofilelist{#1}
  \IfFileExists{#1}{}{\typeout{No file #1.}}
}
\makeatother

\usepackage[,colorlinks,urlcolor=blue,citecolor=black]{hyperref}
\usepackage[normalem]{ulem}

\setuptodonotes{inline}
\usepackage{isomath}
\usepackage{amsmath}
\usepackage{amssymb}
\usepackage{amscd}
\usepackage{amsfonts}

\newcommand{\parderivsec}[2]{\frac{\partial^2 #1}{{\partial #2}^2}}

\newcommand{\parderiv}[2]{\frac{\partial #1}{\partial #2}}

\usepackage{graphicx}% Include figure files
\usepackage{dcolumn}% Align table columns on decimal point
\usepackage{bm}% bold math
\usepackage{color}
\usepackage{xcolor}
\usepackage{epstopdf}
\usepackage{gensymb}
\usepackage{ulem}
\usepackage{float}
\usepackage{xr}
\usepackage{sidecap}
\usepackage{mathrsfs}
\usepackage{amsmath}
\usepackage{amsthm}
\usepackage{enumerate}
\usepackage{soul}
\usepackage{amssymb}
\usepackage{todonotes}
\usepackage{verbatim}

\usepackage{dcolumn}% Align table columns on decimal point
\usepackage{circuitikz}
\usepackage{csquotes}
\usepackage{svg}
\usepackage[symbol]{footmisc}
\usepackage{orcidlink}
\usepackage{verbatim}

\newcommand{\etal}{\textit{et al.}}

\makeatletter
\def\@email#1#2{%
 \endgroup
 \patchcmd{\titleblock@produce}
  {\frontmatter@RRAPformat}
  {\frontmatter@RRAPformat{\produce@RRAP{*#1\href{mailto:#2}{#2}}}\frontmatter@RRAPformat}
  {}{}
}%
\makeatother

\begin{document}

\preprint{APS/123-QED}

\title[Temporal Metasurfaces]{Wave-Freezing and other Phenomena in Temporal Metasurfaces driven by Nonlocal Interactions} 
% Force line breaks with \\
\author{Kshiteej J. Deshmukh\orcidlink{0000-0002-6825-4280}\thanks{*}}
\email{kjdeshmu@math.utah.edu}
\affiliation{Department of Mathematics, University of Utah, Salt Lake City, Utah 84112, U.S.A}

% \email{milton@math.utah.edu}
% \affiliation{ Department of mathematics, University of Utah, Salt Lake City, Utah, U.S.A%\\This line break forced with \textbackslash\textbackslash
% }

% \homepage{http://www.Second.institution.edu/~Charlie.Author.}
%\affiliation{%
%Second institution and/or address%\\This line break forced% with \\
%}%

\date{\today}% It is always \today, today,
%TC:ignore
\begin{abstract}
Space-time metamaterials, or materials with properties changing in space and time, have gained a wide-spread interest due to their exotic properties. 
In this Letter, we propose a novel temporal metasurface of phononic crystals in one and two-dimensions, that combines the use of nonlocal interactions in phononic crystals to customize dispersion relations, and the use of temporal interfaces to transition from a local material to a nonlocal material and vice-versa, to achieve the interesting phenomenon of \textit{wave-freezing}, where the entire propagating wave is stopped without diffusing or spreading.
The phononic crystals are modeled using spring-mass systems and we use finite difference calculations to present our numerical results.
We also demonstrate other effects observed in such temporal metasurfaces, such as time-reversed waves, and anomalous temporal refraction.
\end{abstract}
%TC:endignore
\maketitle
% \todo{-- Add references for the video.\\
% -- change first paragraph\\
% -- think of metasurfaces\\
% -- add time instants to figures\\
% -- check length}
% \begin{quotation}
% The ``lead paragraph'' is encapsulated with the \LaTeX\ 
% \verb+quotation+ environment and is formatted as a single paragraph before the first section heading. 
% (The \verb+quotation+ environment reverts to its usual meaning after the first sectioning command.) 
% Note that numbered references are allowed in the lead paragraph.
% %
% The lead paragraph will only be found in an article being prepared for the journal \textit{Chaos}.
% \end{quotation}

%%%%%%%%%%%%%%%%%%%%%%%%%%%%%
%%%%%%%%%%%%%%%%%%%%%%%%%%%%%
%%%%%%%%%%%%%%%%%%%%%%%%%%%%%
%%%%%%%%%%%%%%%%%%%%%%%%%%%%%

%\section{Introduction:\protect\\ Space-time Metamaterials }\label{sec:Introduction}
Space-time metamaterials or temporal metasurfaces, have physical parameters or material properties varying in time as well as space\cite{lurie2007introduction,caloz2019spacetime,Caloz2020}, providing with even richer physics than metamaterials where properties are only spatially varying, and have resulted in the discovery of exotic properties like space-time mirrors\cite{Bacot2016,moussa2023observation}, chromatic birefringence\cite{Akbarzadeh2018}, space-time cloaking\cite{mccall2010spacetime}, and photonic-time-crystals (PTCs)\cite{zeng2017photonic}.  
They have gained huge popularity in the applications requiring manipulation of  electromagnetic, photonic, acoustic, and elastic wave propagation\cite{fleury2014negative,huidobro2019fresnel,trainiti2019time,pacheco2020temporal,kort2021space}, thus,  attracting the interest of several physicists and engineers. 
%For a detailed review of this topic, the reader is directed to the review articles by Caloz-Deck-Leger\cite{caloz2019spacetime,Caloz2020}, and the book by K. Lurie\cite{lurie2007introduction}.
The time modulation of properties can be done either smoothly (continuously) or abruptly (discontinuously), in which case it is called a time-interface. 
An important distinction between a time-interface and a spatial interface is that the energy of a propagating wave is conserved when it encounters a spatial interface (frequency remains the same), whereas at a time-interface the wave energy typically either increases or decreases, but momentum is conserved (wavevector remains the same). 
In a remarkable experiment with surface water waves by Bacot \etal \cite{Bacot2016}, the effect of such time-interfaces on wave propagation was shown, in which water waves propagating on the surface of a water body split into forward propagating waves and backward propagating waves or time-reversed waves when a vertical jolt was given to the water body. 
These time-reversed waves converged to recreate the image of the initial source or disturbance.

In this Letter, we propose temporal metasurfaces made up of phononic crystals featuring time-interfaces which introduce tailored nonlocal interactions that enable extreme manipulation of the propagating wave, particularly, focusing on the crucial application of \textit{wave-freezing}, wherein a propagating wave is stopped at a location in space without diffusing or spreading. 
The velocity of a propagating wave packet is determined by the slope ($\partial\omega(\kappa)/\partial\kappa$) of the material's dispersion ($\omega$ vs. $\kappa$) curves, where $\omega$ is the frequency and $\kappa$ is the wavevector.
In elastic and acoustic materials, researchers are interested in achieving waves with zero group-velocity $\partial\omega/\partial \kappa=0$, the so-called zero-group-velocity(ZGV) modes, which are important in the applications of non-destructive testing and quantitative characterization of structures\cite{mora2022nonlinear,kiefer2023computing,samak2024evidence}.
Such modes are characterized by the critical points of the dispersion relation.
%\textit{Wave-freezing} is of great relevance and importance to applications that involve slow light or stopping of light, towards which there have been several efforts\cite{liu2001observation,figotin2003oblique,baba2008slow,khurgin2010slow,figotin2011slow,yamilov2023anderson}. 
In photonics and optics, several mechanisms have been proposed to obtain ZGV modes at exceptional points and band edge, that are crucial in the observation of slow-light or stopping of light\cite{liu2001observation,figotin2003oblique,baba2008slow,khurgin2010slow,figotin2011slow,goldzak2018light,yamilov2023anderson}. 
Although the existence of such nonpropagating modes is well known, an important question is how can one stop the propagating wave at a given location for as long a duration as desired without any scattering?
For photonic crystals, Figotin \etal\cite{figotin2001nonreciprocal,figotin2003oblique,figotin2006slow} proposed the nonreciprocal frozen mode regime to address this issue in certain cases.
Using anisotropic layered media, a stationary inflection point $\partial^2\omega / \partial\kappa^2=0$ in the dispersion relation was achieved \cite{ballato2005frozen}, which is even more desirable for wave-freezing as the value of $\partial^2\omega(\kappa) / \partial\kappa^2$, governs the wave-packet diffusion dynamics.
In our work, we present a simple mechanism to obtain on-demand freezing of the entire propagating wave in one-dimensional (1-D) and two-dimensional (2-D) phononic temporal metasurfaces for a broad range of frequencies and wavevectors inside the Brillouin Zone (BZ).
We also highlight the phenomena of time-mirrors\cite{Bacot2016,moussa2023observation} and anomalous temporal refraction that occur when such time-interfaces, where the strength of nonlocal interactions changes with respect to time, are considered.

Central to our work is the result that by using nonlocal interactions up to $N^{th}$ nearest-neighbors, the dispersion relation for the first-band of a phononic crystal takes the form of a truncated Fourier series, as was first shown by Chen-Kadic-Wegener\cite{chen2021roton}, and can have critical points inside the first BZ.
Subsequently, it was shown that one can obtain obtain any desired dispersion curve as the first band of a 2-D acoustic or elastic metamaterial\cite{wang2022nonlocal}.
In\cite{kazemi2023drawing}, an inverse design methodology was proposed that allows customization of the first two dispersion bands of a nonlocal phononic crystal in 1-D by designing the strength of nonlocal interactions.

We first consider 1-D phononic crystals with nearest and up to $N^{th}$ nearest neighbor interactions modeled by a discrete spring-mass system with the mass displacements along the 1-D. 
%Later, we introduce two-dimensional (2-d) nonlocal phononic crystals with nonlocal interactions in both directions and the mass displacements are considered in the third, out-of-plane, direction.
The equations of motion are solved numerically using the finite-difference method. 
All values in this Letter are to be considered in arb. units.

\begin{figure*}[t]
\centering
        \includegraphics[width=0.9\linewidth]{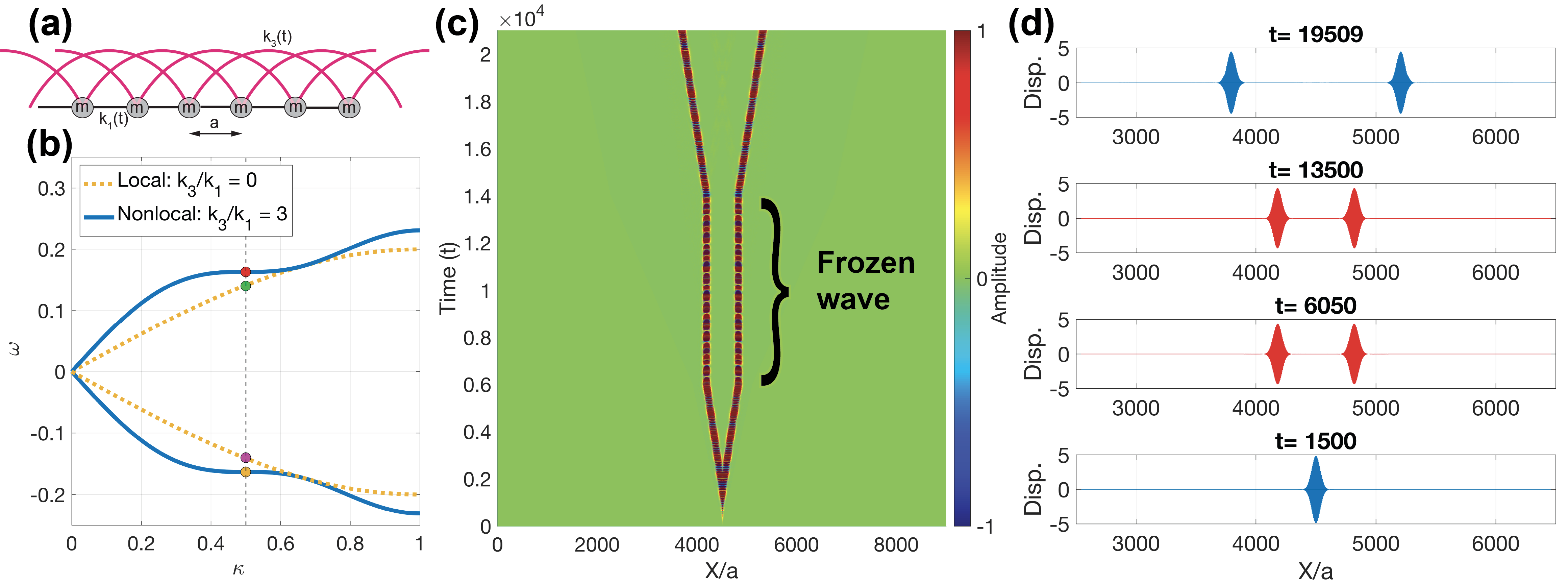} 
        \caption{Wave-freezing in 1-D: (a) Schematic of the spring-mass system with nearest and third-nearest neighbor interactions. 
        (b) Dispersion curves, $\omega$ vs. $\kappa$, for the spring-mass system with $N=1$ (local interactions only), i.e., $k_1=0.01, k_3=0$ (dotted yellow curve), and with $N=3$, i.e., $k_1=0.01, k_2=0, k_3=k_1/3$ (solid blue curve). 
        The green marker shows the frequency-wavevector pair of the propagating wave before the time-interface, red and yellow markers show the frequency-wavevector pairs of the frozen-wave, and pink marker corresponds to the backward propagating wave after the $2^{nd}$ time-interface.
        (c) X-t plot of the wave displacement: The vertical red lines in the plot show that the left and right propagating waves are frozen in space for a finite duration of time as long as the nonlocal interactions are present. 
        After the $2^{nd}$ time-interface, a dominant forward propagating wave continues traveling in the original system as before, with negligible amplitude of the backward propagating wave.
        (d) Displacement snapshots of wave propagation at different instants of time (frozen waves shown in red).
        }
        \label{fig:1d_freezing}
\end{figure*}
Consider a 1-D spring-mass chain with the lattice constant denoted by $a$, and the neighboring interactions of the masses modeled by linear springs with stiffness values that can change with respect to time $(t)$.
All masses are considered identical and are denoted by $m$. 
For all examples considered in this work, we choose $m=1$, and $a=\pi$.
Let $k_1(t)$ denote the spring stiffness values for the nearest-neighbor or local interactions, and $k_2(t), k_3(t),\dots, k_N(t)$ denote the spring stiffness values for the second, third, \dots, and $N^{th}$ nearest-neighbor interactions, respectively.  
%The displacement of the $i^{th}$ mass is denoted by $u_i(t)$. 
A schematic of such a spring-mass system with nonlocal interactions in 1-D is shown in Fig. \ref{fig:1d_freezing}\textcolor{red}{(a)}. 
% \begin{figure*}[t]
%         \includegraphics[width=0.8\linewidth]{Figure_3modes_main_2.png} 
%         \caption{Image reconstruction from time-reversed waves in nonlocal materials. 
%         (a) $\omega$ vs. $\kappa$, dispersion curves for the local(yellow dotted curve) and nonlocal(solid blue curve) materials. 
%         The wave packet propagating at $\omega=0.25$ in the nonlocal material splits into 3 modes with the same $\omega$ but different $\kappa$ values (shown by the green circle markers).
%         At the $1^{st}$ time-interface, the nonlocal interactions vanish, and only local interactions are present. 
%         The 3 modes after the $1^{st}$ time-interface, transition to the 3 modes shown by red and blue circle markers.
%         (b) X-T plot of the displacement amplitude showing image reconstruction from 3 backward propagating modes after the $2^{nd}$ time-interface. 
%         (c) Snapshots of the displacement at different time instants. 
%         The red curves show the displacements after the $2^{nd}$ time-interface where we have 3 forward propagating modes and 3 backward propagating modes.
%         At $t=19298$, the 3 backward propagating modes reconstruct to form the image of the source.
%         }
%         \label{fig:1d_3modes}
% \end{figure*}
The equation of motion that governs the displacement $u_i(t)$ of the $i^{th}$ mass is given as, 
\begin{equation}\label{EoM}
    m \Ddot{u}_i = \sum_{n=1}^{N} k_i (u_{i+n} - 2u_i + u_{i-n} ) +f(t),
\end{equation}
where, the dot denotes derivative with respect to time $t$.
Using the Bloch form\cite{bloch1929quantenmechanik} of the wave solution, we obtain the following form of the dispersion relation:
\begin{equation}\label{dispersion_relation_1d}
    \omega^2 = \frac{4}{m}\sum_{n=1}^{N}k_n \sin^2{\Big(\frac{n\kappa a}{2}\Big)}.
\end{equation} 
%Using the analysis in \cite{kazemi2023drawing}, we then compute the values ${k_1, k_2, \dots, k_N}$ to introduce tailored nonlocal interactions to obtain any desired dispersion relation (upto an approximation by the truncated Fourier series).
At time $t=0$, we begin by considering a spring-mass chain with $N=1$, i.e., with $k_1 = 0.01$, and  $k_2=k_3=0=\dots=k_N = 0$.
With these values of $k_n$ in \eqref{dispersion_relation_1d}, we recover the well-known dispersion relation for a 1-D spring mass chain with nearest neighbor interactions only (shown by the dotted yellow curve in Fig. \ref{fig:1d_freezing}\textcolor{red}{(b)}). 
The center mass of the spring-mass system is excited by a constant frequency forcing given as, $f(t) = \exp{\big(-(t-\mu)^2 / \tau\big)} \cos(\omega_0 t)$, where, $\tau$ and $\mu$ are parameters of the Gaussian envelop of the forcing function, and $\omega_0=0.15$ is the carrier frequency.
The forcing produces right propagating and left propagating wave packets with the frequency-wavevector pair $(\omega_0, \kappa_0) = (0.15, 0.5)$ (corresponding to the green marker in Fig. \ref{fig:1d_freezing}\textcolor{red}{(b)}).
Figure \ref{fig:1d_freezing}\textcolor{red}{(c)} shows the $X-t$ plot of the displacement with the color representing the amplitude of the displacements. 
The $1^{st}$ time-interface is modeled at $t=6050$, by introducing tailored nonlocal interactions with $k_1=0.01, k_2=0, k_3 = k_1/3$, and transitioning to a $N=3$ system.
The solid blue curve in Fig. \ref{fig:1d_freezing}\textcolor{red}{(b)} shows the dispersion relation of this nonlocal system.
At the time-interface, the wavevector of the propagating waves is preserved and is the same as $\kappa_0$, while $\omega$ changes (red and yellow markers in Fig. \ref{fig:1d_freezing}\textcolor{red}{(b)}).
%Here, our goal is to "freeze" the wave, i.e., to stop the propagation of wave. 
The nonlocal interactions were so tailored, that the dispersion relation of the resulting system has a vanishing second partial derivative of frequency with respect to wavevector, i.e., $\parderivsec{\omega}{\kappa}\Big\vert_{\kappa=\kappa_0}=0$ at the wavevector $\kappa_0$ of the propagating wave. 
This means that not only the group-velocity of the wave is $0$, but also that there is no diffusion or spreading of the wave packet (see \cite{remoissenet2013waves,kazemi2023drawing} for more detailed discussion of wave-packet diffusion).
As a result, there is no splitting of the wave-packet after the first time-interface, and the propagation of the entire Gaussian wave-packet is stopped, i.e., the wave is frozen at a location in space for as long as desired without diffusing. 
The frozen waves are seen in Fig. \ref{fig:1d_freezing}\textcolor{red}{(c)},  as the two vertical straight band portions.
At $t=13500$, we implement the second time-interface, at which the nonlocal interactions are reduced to $0$, and we get back to the original spring-mass chain with $N=1$ and $k_1=0.01$. 
At the $2^{nd}$ time-interface, the wave packet splits into a forward propagating wave and a backward propagating wave (corresponding to the green and pink markers in Fig. \ref{fig:1d_freezing}\textcolor{red}{(b)}, respectively) with the frequency-wavevector pair $(\pm\omega_0, \kappa_0)$.
We observe that the backward propagating wave is of negligible amplitude, and hence can't be seen in the $X-t$ plot of Fig. \ref{fig:1d_freezing}\textcolor{red}{(c)}, while the forward propagating wave dominantly carries the energy of the system.
% The amplitude of the backward propagating wave depends on the instant of time at which one applies the $2^{nd}$ time-interface.
Negligible to small amplitudes of the backward propagating wave can be observed depending on the instant at which one applies the $2^{nd}$ time-interface (examples provided in the Supplementary Information document (SI)\cite{SI}).
Figure \ref{fig:1d_freezing}\textcolor{red}{(d)} shows snapshots of the wave displacement, and we see that from $t=6050$ to $t=13500$ (instants of first and second time-interfaces, respectively), the wave envelop remains frozen (shown in red) at the location, and starts propagating again after the second time-interface.

It should be noted that, wave-freezing is not limited to a specific $(\omega, \kappa)$ pair. 
Taking advantage of the Fourier series representation of the dispersion relation \eqref{dispersion_relation_1d}, one can easily (analytically) obtain tailored ${k_1, k_2, \dots, k_N}$ values so that a flat band (with $\parderivsec{\omega}{\kappa} = 0$) can be obtained for a sufficiently broad range of $\kappa$ values (for e.g., see Fig. 2(a) in \cite{kazemi2023drawing}).

%%%%%%%%%%%%%%%%%%%%%%%%%%
%%%%%%%%%%%%%%%%%%%%%%%%%%
%%%%%%%%%%%%%%%%%%%%%%%%%%
\begin{figure}[t]
        \includegraphics[width=\linewidth]{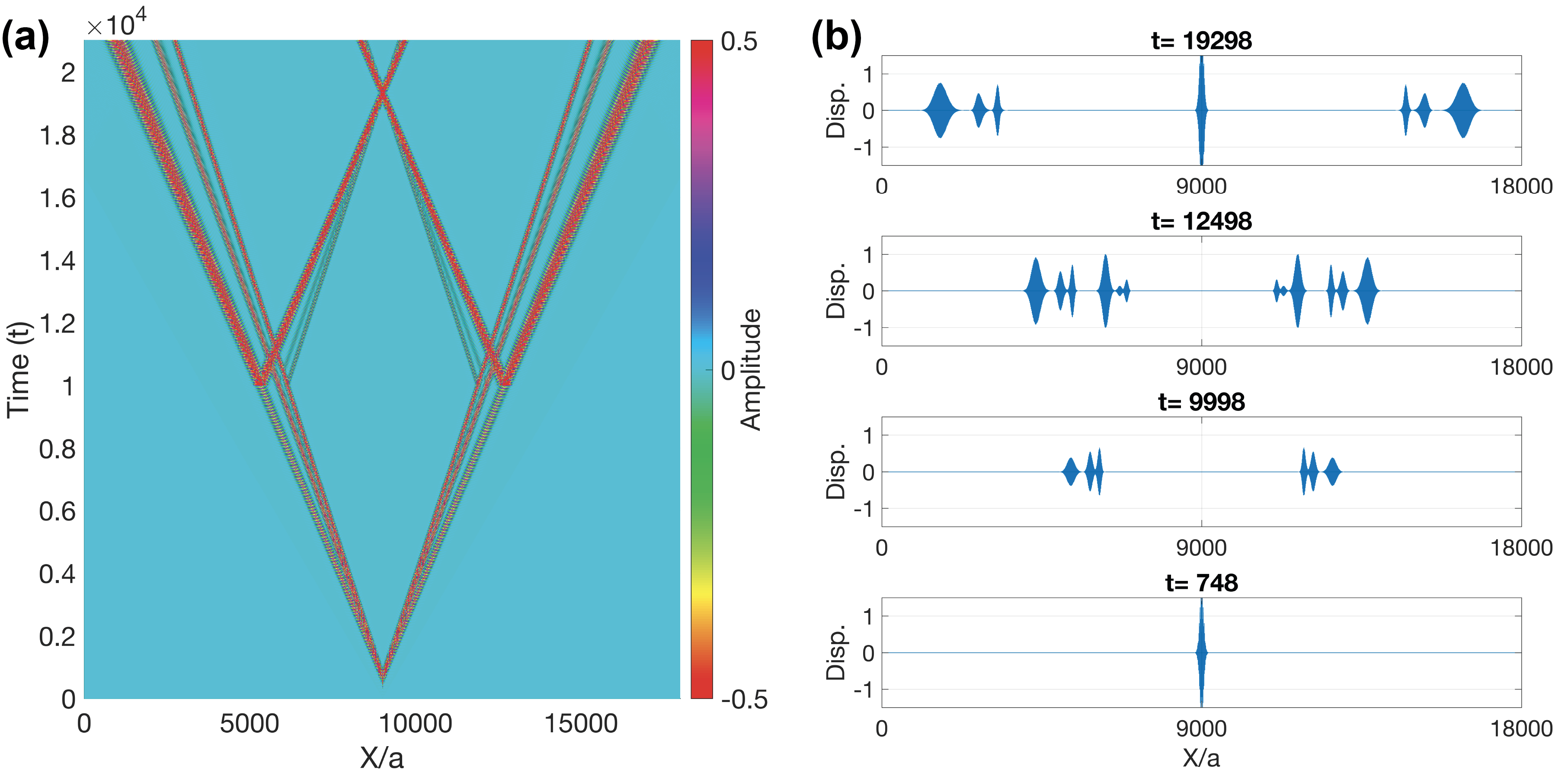} 
        \caption{Image reconstruction from time-reversed waves in nonlocal materials: 
        (a) X-t plot of the displacement amplitude showing image reconstruction from three backward propagating modes. 
        (b) Snapshots of the wave displacement profile at different time instants.
        At $t=19298$, the 3 backward propagating modes reconstruct to form the image of the source.
        }
        \label{fig:1d_3modes}
\end{figure}
Typically, time modulation of material properties in the presence of a propagating wave propagating requires energy to be added to or removed from the system (except in some energy conserving temporal metasurfaces\cite{deshmukh2022energy}).
For wave-freezing in 1-D, we present the energy exchange associated with the introduction and removal of nonlocal interactions in Fig. \textcolor{red}{S1} of the SI\cite{SI}. 
We observe that at the $1^{st}$ time-interface as the nonlocal spring with stiffness $k_3$ is introduced in the system, the energy ($E$) of the system increases by an amount $\Delta E_1$, and the wave is frozen. 
At the $2^{nd}$ time-interface, we set $k_3=0$, and energy of the spring-mass system decreases by an amount $\Delta E_2$. 
Interestingly, when $\Delta E_1 > \Delta E_2$, a small amplitude of the backward propagating wave is observed, and when $\Delta E_1 \approx  \Delta E_2$ almost negligible amplitude of the backward propagating wave is observed.

%%%%%%%%%%%%%%%
%%%%%%%%%%%%%%%
%%%%%%%%%%%%%%%
%%%%%%%%%%%%%%%
Next, we demonstrate image reconstruction from the time-reversal of waves in nonlocal 1-D temporal metasurfaces.
Consider a 1-D spring-mass chain that has nonlocal interactions to begin with, we choose, $N=3$, with $k_1 = 0.01, k_2 = 0, k_3/k_1=3$. 
The dispersion relation of the nonlocal system is shown in Fig. \textcolor{red}{S2} by the solid blue curve. 
When the center mass of the spring mass chain is forced to oscillate at $\omega=0.25$, it produces three modes traveling to the right and three modes traveling to the left (see, SI for more details). 
% The 3 right traveling modes are represented by the green circle markers in Fig. \ref{fig:1d_3modes}\textcolor{red}{(a)}, which show that they have the same $\omega$, but different $\kappa$ values.
At $t=9998$, we model the $1^{st}$ time-interface, where the spring stiffness values of the nonlocal springs are reduced to $0$, i.e., $k_3=0$, and then after time $\Delta t = 10$ we model the $2^{nd}$ time-interface where original stiffness values of the springs are restored,i.e., $k_1 = 0.01, k_2=0, k_3/k_1=3$.
The $X-t$ plot of the displacement amplitude and the snapshots of the displacement profiles are shown in Figs. \ref{fig:1d_3modes}\textcolor{red}{(a)-(b)},  respectively.
% When the propagating modes encounter the $1^{st}$ time-interface, they are propagating in a material that has local interactions only (dispersion relation shown by yellow dotted curve in Fig. \ref{fig:1d_3modes}\textcolor{red}{(a)}), and the 3 propagating modes now transition to modes which have different $\omega$ and $\kappa$ values (shown by red and blue markers in Fig. \ref{fig:1d_3modes}\textcolor{red}{(a)}).
After the $2^{nd}$ time-interface, each of the 3 modes splits into a forward propagating and backward propagating wave in nonlocal material (shown by the displacement snapshot at $t=12498$ in Fig. \ref{fig:1d_3modes}\textcolor{red}{(b)}).
The 3 backward propagating modes, reconstruct to form an image of the initial wave packet (see snapshot at $t=19298$ in Fig. \ref{fig:1d_3modes}\textcolor{red}{(b)}). 
The case with just a single interface is shown in Fig. \textcolor{red}{S3} of the SI, where multiple images of the source are obtained from time-reversed waves consisting of different frequencies. 
%\textcolor{blue}{Such temporal metasurfaces can be useful in the applications involving frequency multiplexing and frequency splitting}.

%%%%%%%%%%%%%%%
%%%%%%%%%%%%%%%
%%%%%%%%%%%%%%% Wave-freezing in 2-D:
%%%%%%%%%%%%%%%
\begin{figure}[tb!]
        \centering
        \includegraphics[width=0.96\linewidth]{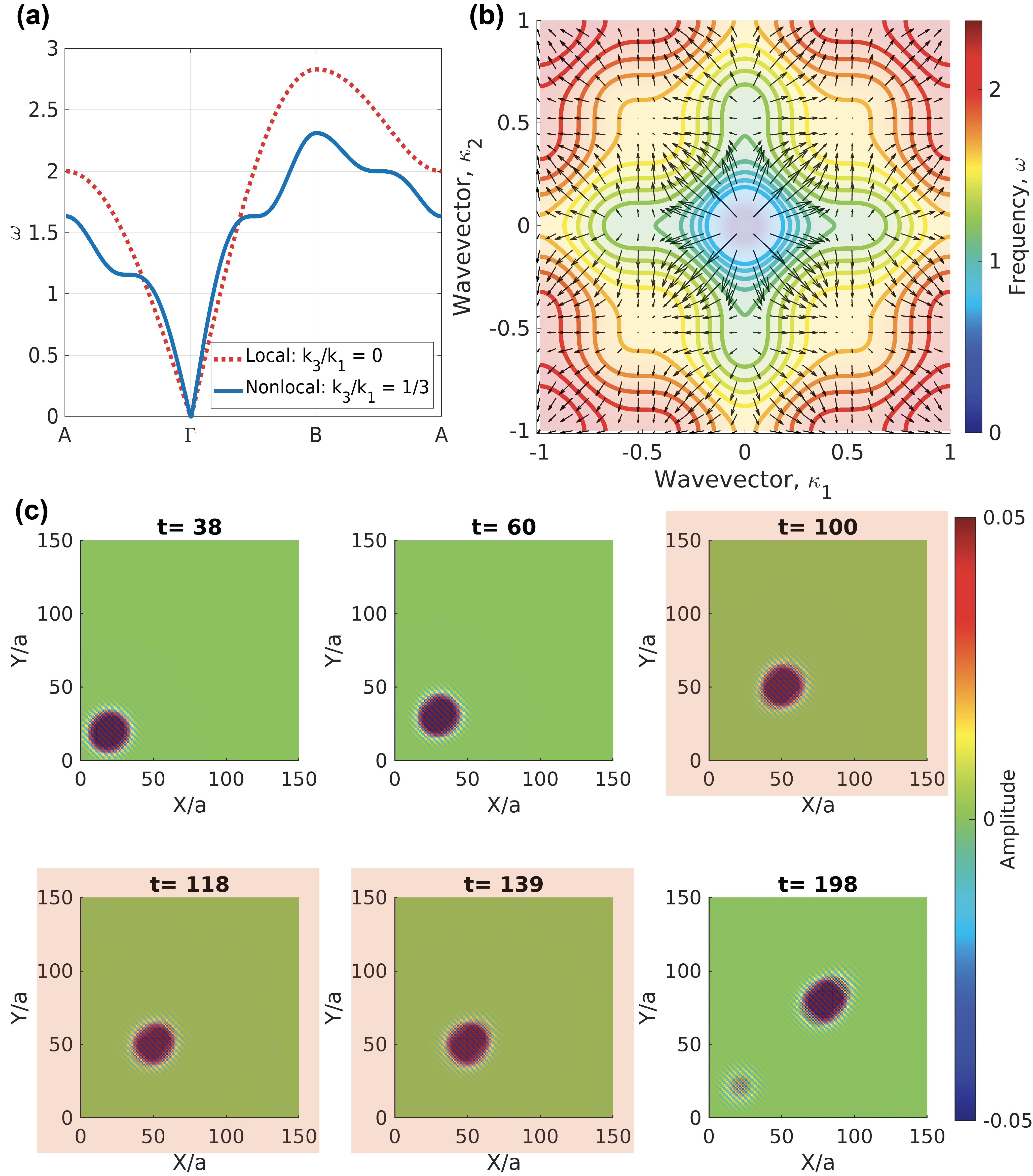} 
        \caption{Wave-freezing in 2-D: 
        (a) Dispersion curves along the BZ boundary for the local (dotted red curve) and nonlocal (solid blue curve) systems.
        (b) Iso-frequency contours of the nonlocal system. 
        Colorbar represents the $\omega$ values.
        The black arrows represent the group-velocity vector-field for different $(\kappa_1, \kappa_2)$ values. 
        (c) The displacement amplitude of the Gaussian wave-packet is shown as it propagates along a line at an angle of $45^\circ$ with the $X$-axis. 
        At $t=100$, the local system transitions to a tailored nonlocal system, and the wave-packet remains frozen at the same-location as long as the nonlocal interactions are present, i.e., from $t=100$ to $t=139$. 
        The frozen wave snapshots are highlighted in red color.
        At the $2^{nd}$ time-interface when the system becomes local again, the wave packet splits into a (dominantly) forward propagating wave, and a backward propagating wave of much smaller amplitude as shown by the displacement snapshot at $t=198$.  
        }
        \label{fig:Wave_freezing_2D}
\end{figure}
The wave-freezing  mechanism proposed above is robust even in 2-D which we demonstrate below. 
In 2-D, we consider masses with out-of-plane displacements (scalar wave propagation) having nonlocal interactions in both the $X$ and $Y$ directions.
Let $N$ and $M$ denote the number of nearest neighbors in the $X$ and $Y$ directions, respectively.
Let ${k_{x_1}(t), k_{x_2}(t), \dots, k_{x_N}(t) }$ denote the spring stiffness values modeling the nonlocal interactions of the $i^{th}$ nearest neighbor in the $X$ direction,  and let ${k_{y_1}(t), k_{y_2}(t), \dots, k_{y_M}(t) }$  denote the spring stiffness values modeling the nonlocal interactions of the $j^{th}$ nearest neighbor in the $Y$ direction. 
The unit cell of such a 2-D metasurface is shown in Fig. \textcolor{red}{S4} of the SI. 
For finite  values of $N$ and $M$, the dispersion relation of such a nonlocal spring-mass system again turns out to be of the form of a truncated Fourier series\cite{wang2022nonlocal}, which is given by, 
\begin{equation}\label{2d_DR}
    \omega^2(\kappa_1, \kappa_2) = \frac{-2}{m}
                                \Bigg(
                                    \sum_{i=0}^N k_{x_i} \cos{(i \kappa_1 a)}
                                    +
                                    \sum_{j=0}^M k_{y_j} \cos{(j \kappa_2 a)}
                                    \Bigg),
\end{equation}
where, $(\kappa_1, \kappa_2)$ are wavevectors in the $X$ and $Y$ directions, respectively (see SI for more information).
Wang-Chen \etal\cite{wang2022nonlocal} used nonlocal interactions with $N=M$, to engineer dispersion relations in acoustic and mechanical metamaterials.
We begin by considering a 2-D spring-mass system with local interactions only, i.e., $N=M=1$ and $k_{x_1}=k_{y_1} \equiv k_1= 1$. 
At $t=0$, the masses are given an initial displacement in the form of a Gaussian wave-packet given by the expression, $e^{-(x^2+y^2)/1000}\sin{(0.5 x)}\sin{(0.5 y)}$, resulting in a wave propagating from the left bottom corner of the 2-D system at an angle of $45^\circ$ with the $X$ direction, as shown in the displacement amplitude snapshots at different instants of time in Fig. \ref{fig:Wave_freezing_2D}\textcolor{red}{(c)}.
At the $1^{st}$ time-interface at $t=100$, we introduce nonlocal springs with stiffness values given by $k_{x_3} = k_{y_3}\equiv k_3 = k_1/3$, thus, transitioning from a completely local to a nonlocal system with $N=M=3$.
The dispersion relations of the local and nonlocal systems along the BZ boundary are shown in Fig. \ref{fig:Wave_freezing_2D}\textcolor{red}{(a)} by the dotted red curve and solid blue curve, respectively (see SI for details on the BZ boundary).
Fig. \ref{fig:Wave_freezing_2D}\textcolor{red}{(b)} shows the iso-frequency contours of the nonlocal system as a function of $(\kappa_1, \kappa_2)$, with the colorbar denoting $\omega$ values. 
The black arrows show the group-velocity vector field, i.e., $\Big[\parderiv{\omega}{\kappa_1}, \parderiv{\omega}{\kappa_2}\Big]$, and we observe some regions with vanishing group-velocities. 
In fact, for the given problem, the values of $k_{x_3}, k_{y_3}$ are so chosen that, again we not only have zero group-velocity, but also the Hessian of $\omega(\kappa_1, \kappa_2)$ is zero at $(\kappa_1, \kappa_2) = (0.5,0.5)$, i.e.,
\begin{equation}
    %\Big[\parderiv{\omega}{\kappa_1}, \parderiv{\omega}{\kappa_2}\Big]\Big\vert_{(\kappa_1, \kappa_2) = (0.5,0.5)} =\bf0, 
    %\quad 
    \begin{bmatrix}
    \parderivsec{\omega}{\kappa_1}& \frac{\partial^2\omega}{\partial\kappa_1\partial\kappa_2}
    \\
    \frac{\partial^2\omega}{\partial\kappa_1\partial\kappa_2} & \parderivsec{\omega}{\kappa_2}
    \end{bmatrix}_{(\kappa_1, \kappa_2) = (0.5,0.5)}
     =
    \bf0.
\end{equation}
This is the point seen as the flat portion of the solid curve in Fig. \ref{fig:Wave_freezing_2D}\textcolor{red}{(a)} in the region $\Gamma-B$.
At the time-interface, as we transition to this nonlocal system the wavevectors, $(\kappa_1, \kappa_2) = (0.5,0.5)$,  are conserved. 
As a result, the entire Gaussian wave-packet is now frozen at the same location for as long as the nonlocal interactions are present without diffusing.
This frozen wave is shown by the displacement snapshots highlighted in red color in Fig. \ref{fig:Wave_freezing_2D}\textcolor{red}{(c)}.
After sometime, we change $k_{x_3} = k_{y_3}\equiv k_3 = 0$, at which the frozen wave encounters a $2^{nd}$ time-interface and generates in the local system, a forward propagating wave traveling in the same direction as before, and a backward propagating wave (seen at $t=198$ in Fig. \ref{fig:Wave_freezing_2D}\textcolor{red}{(c)}).
The forward propagating wave carries the dominant part of the energy, and the backward propagating wave is of a relatively small amplitude.

\begin{figure}[t]
        \includegraphics[width=\linewidth]{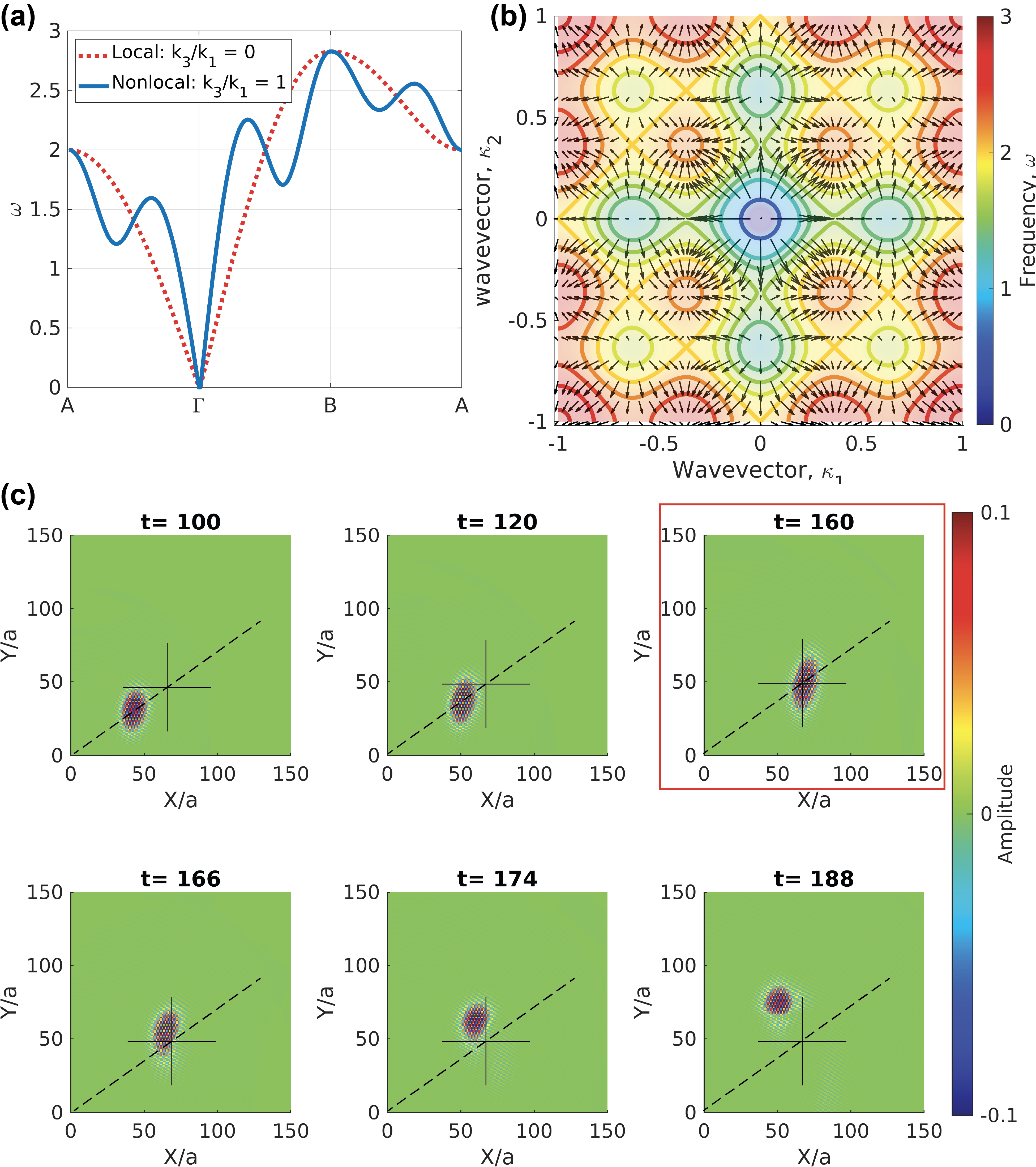} 
        \caption{Anomalous temporal refraction:
        (a) Dispersion curves along the BZ boundary for the local (dotted red curve) and nonlocal (solid blue curve) systems.
        (b) Iso-frequency contours of the nonlocal system. 
        Colorbar represents $\omega$ values.
        The black arrows represent the group-velocity  vector-field for different $(\kappa_1, \kappa_2)$ values. 
        Regions with one or both components of group-velocity being negative can be observed.
        (c) The displacement amplitude of the Gaussian wave-packet is shown as it propagates initially along the black dotted line, and the black cross-hairs aid in visualizing the direction of propagation of the wave-packet.
        At $t=160$ (highlighted by the red rectangle), nonlocal springs  with stiffness values of ${k_x}_3={k_y}_3 = {k_x}_1$ are introduced. 
        We observe anomalous temporal refraction as the wave packet splits into a forward propagating wave packet that propagates into the second-quadrant and a backward propagating wave that propagates into the fourth-quadrant.
        }
        \label{fig:negative_refraction}
\end{figure}

%%%%%%%%%%%%%%%
%%%%%%%%%%%%%%%
%%%%%%%%%%%%%%% ANOMALOUS REFRACTION
%%%%%%%%%%%%%%%
In Fig. \ref{fig:negative_refraction}, we show an interesting case of anomalous refraction at the temporal interface as one transitions from a local system with $k_{x_1} = k_{y_1}\equiv k_1 = 1$ to a nonlocal system with $N=M=3$, by introducing nonlocal springs with $k_{x_3} = k_{y_3}\equiv k_3 = k_1$. 
The dispersion curves for the local (dotted red) and nonlocal (solid blue) systems along the BZ boundary are shown in Fig. \ref{fig:negative_refraction}\textcolor{red}{(a)}. 
Figure \ref{fig:negative_refraction}\textcolor{red}{(b)} shows the iso-frequency contours of the nonlocal system as a function of $(\kappa_1, \kappa_2)$, with the colorbar denoting $\omega$ values. 
The group-velocity vector field represented by the black arrows shows the regions with negative group-velocities as well as vanishing group-velocities.
Similar representation of the group-velocity field for the local metasurface is shown in Fig. \textcolor{red}{S5} of the SI. 
Figure \ref{fig:negative_refraction}\textcolor{red}{(c)} shows the snapshots of the displacement amplitude of the propagating wave.
We consider an initial Gaussian wave-packet given by the expression $ e^{-(x^2+y^2)/500}\sin{(0.5 x)}\sin{(0.75 y)}$, propagating in the local medium along the dashed black line (the cross-hairs help in visualizing the direction of propagation of the wave-packet.).
At $t=160$, the local system transitions to the nonlocal system mentioned above, which results into a forward propagating wave-packet traveling in the second-quadrant and a backward propagating wave-packet traveling in the fourth-quadrant.
If the temporal metasurface had local interactions only, then the forward and backward propagating wave packets would travel  in the first and third quadrants only, due to the group-velocity components being both positive or both negative. 
Nonlocal interactions make it possible to have either or both components of group-velocity to be negative, as can be seen in Fig. \ref{fig:negative_refraction}\textcolor{red}{(b)}, thus resulting in this anomalous temporal refraction.
The flexibility offered by such temporal metasurfaces make them a great candidate for the applications of temporal aiming\cite{pacheco2020temporal} and beam steering.
%(it is to be noted that nonlocal interactions are not the only way to observe this effect). 

%%%%%%%%%%%%%%%%%%%%%%%%%%
%%%%%%%%%%%%%%%%%%%%%%%%%%
%%%%%%%%%%%%%%%%%%%%%%%%%%
%%%%%%%%%%%%%%%%%%%%%%%%%%
%%%%%%%%%%%%%%%%%%%%%%%%%%
%%%%%%%%%%%%%%%%%%%%%%%%%%
%\section{\label{sec:Conclusion} Conclusion}
In conclusion, we have proposed novel temporal metasurfaces that employ time-interfaces introducing engineered nonlocal interactions to achieve extreme wave manipulation, especially focusing on the phenomenon of \textit{wave-freezing}, that is of wide-spread interest.
The idea is demonstrated in 1-D and 2-D by using a discrete spring-mass system, and can be easily extended to three-dimensions.
Although the examples considered in this work deal with waves propagating at a given frequency-wavevector value, but in principle, the proposed mechanism is applicable for a broad range of frequency and wavevector values inside the first BZ, except near the origin.
Nonlocal acoustic and mechanical metamaterials have been realized in some recent notable works\cite{chen2021roton, wang2022nonlocal}. 
At present, it is not clear and requires further thought on how one would change the strength of nonlocal interactions in such materials to realize the time-interfaces discussed in this work.
Alternatively, it is easier to experimentally realize a transmission line model  consisting of inductors, capacitors, and switches (assuming negligible losses due to resistance) that is exactly analogous to the spring-mass systems considered in this work (see SI\cite{SI} for more discussion and a proposed model).

%%%%%%%%%%%%%%%%%%%%%%%%%%%%%
%%%%%%%%%%%%%%%%%%%%%%%%%%%%%
%%%%%%%%%%%%%%%%%%%%%%%%%%%%%
%%%%%%%%%%%%%%%%%%%%%%%%%%%%%
%TC:ignore
%\section*{Supplementary Information}
%See SI material for additional details related to this work.
 
The author would like to thank Prof. Graeme W. Milton at the University of Utah for his helpful discussions and encouragement in publishing this work.
The author is grateful to the National Science Foundation for support through grant DMS-2107926.  
%\appendix
%%%%%%%%%%%%%%%%%%%%%%%%%%%%%
%%%%%%%%%%%%%%%%%%%%%%%%%%%%%
%%%%%%%%%%%%%%%%%%%%%%%%%%%%%
%%%%%%%%%%%%%%%%%%%%%%%%%%%%%

%%%%%%%%%%%%%%%%%%%%%%%%%%%%%
%%%%%%%%%%%%%%%%%%%%%%%%%%%%%
%%%%%%%%%%%%%%%%%%%%%%%%%%%%%
%%%%%%%%%%%%%%%%%%%%%%%%%%%%%
 
%%%%%%%%%%%%%%%%%%%%%%%%%%%%%
%%%%%%%%%%%%%%%%%%%%%%%%%%%%%
%%%%%%%%%%%%%%%%%%%%%%%%%%%%%
%%%%%%%%%%%%%%%%%%%%%%%%%%%%%
% \section*{Author Declarations}
% \subsection*{Conflict of Interest}
% The authors have no conflicts to disclose.

% \subsection*{Authors' Contribution}
% K. J. Deshmukh did all the work in this article.
%%%%%%%%%%%%%%%%%%%%%%%%%%%%%
%%%%%%%%%%%%%%%%%%%%%%%%%%%%%
%%%%%%%%%%%%%%%%%%%%%%%%%%%%%
%%%%%%%%%%%%%%%%%%%%%%%%%%%%%
% \section*{Data Availability}
% Aside from the data available within the article, any other data are available from the corresponding author upon request. 

\section*{References}

%%%%%%%%%%%%%%%%%%%%%%%%%%%%%
%%%%%%%%%%%%%%%%%%%%%%%%%%%%%
%%%%%%%%%%%%%%%%%%%%%%%%%%%%%
%%%%%%%%%%%%%%%%%%%%%%%%%%%%%
%\nocite{*}
\bibliography{Main}% Produces the bibliography via BibTeX.

\begin{thebibliography}{33}%
\makeatletter
\providecommand \@ifxundefined [1]{%
 \@ifx{#1\undefined}
}%
\providecommand \@ifnum [1]{%
 \ifnum #1\expandafter \@firstoftwo
 \else \expandafter \@secondoftwo
 \fi
}%
\providecommand \@ifx [1]{%
 \ifx #1\expandafter \@firstoftwo
 \else \expandafter \@secondoftwo
 \fi
}%
\providecommand \natexlab [1]{#1}%
\providecommand \enquote  [1]{``#1''}%
\providecommand \bibnamefont  [1]{#1}%
\providecommand \bibfnamefont [1]{#1}%
\providecommand \citenamefont [1]{#1}%
\providecommand \href@noop [0]{\@secondoftwo}%
\providecommand \href [0]{\begingroup \@sanitize@url \@href}%
\providecommand \@href[1]{\@@startlink{#1}\@@href}%
\providecommand \@@href[1]{\endgroup#1\@@endlink}%
\providecommand \@sanitize@url [0]{\catcode `\\12\catcode `\$12\catcode
  `\&12\catcode `\#12\catcode `\^12\catcode `\_12\catcode `\%12\relax}%
\providecommand \@@startlink[1]{}%
\providecommand \@@endlink[0]{}%
\providecommand \url  [0]{\begingroup\@sanitize@url \@url }%
\providecommand \@url [1]{\endgroup\@href {#1}{\urlprefix }}%
\providecommand \urlprefix  [0]{URL }%
\providecommand \Eprint [0]{\href }%
\providecommand \doibase [0]{http://dx.doi.org/}%
\providecommand \selectlanguage [0]{\@gobble}%
\providecommand \bibinfo  [0]{\@secondoftwo}%
\providecommand \bibfield  [0]{\@secondoftwo}%
\providecommand \translation [1]{[#1]}%
\providecommand \BibitemOpen [0]{}%
\providecommand \bibitemStop [0]{}%
\providecommand \bibitemNoStop [0]{.\EOS\space}%
\providecommand \EOS [0]{\spacefactor3000\relax}%
\providecommand \BibitemShut  [1]{\csname bibitem#1\endcsname}%
\let\auto@bib@innerbib\@empty
%</preamble>
\bibitem [{\citenamefont {Lurie}(2007)}]{lurie2007introduction}%
  \BibitemOpen
  \bibfield  {author} {\bibinfo {author} {\bibfnamefont {K.~A.}\ \bibnamefont
  {Lurie}},\ }\href@noop {} {\emph {\bibinfo {title} {An introduction to the
  mathematical theory of dynamic materials}}},\ Vol.~\bibinfo {volume} {15}\
  (\bibinfo  {publisher} {Springer},\ \bibinfo {year} {2007})\BibitemShut
  {NoStop}%
\bibitem [{\citenamefont {Caloz}\ and\ \citenamefont
  {Deck-L{\'e}ger}(2019)}]{caloz2019spacetime}%
  \BibitemOpen
  \bibfield  {author} {\bibinfo {author} {\bibfnamefont {C.}~\bibnamefont
  {Caloz}}\ and\ \bibinfo {author} {\bibfnamefont {Z.-L.}\ \bibnamefont
  {Deck-L{\'e}ger}},\ }\href@noop {} {\bibfield  {journal} {\bibinfo  {journal}
  {IEEE Transactions on Antennas and Propagation}\ }\textbf {\bibinfo {volume}
  {68}},\ \bibinfo {pages} {1569} (\bibinfo {year} {2019})}\BibitemShut
  {NoStop}%
\bibitem [{\citenamefont {Caloz}\ and\ \citenamefont
  {Deck-L{\'{e}}ger}(2020)}]{Caloz2020}%
  \BibitemOpen
  \bibfield  {author} {\bibinfo {author} {\bibfnamefont {C.}~\bibnamefont
  {Caloz}}\ and\ \bibinfo {author} {\bibfnamefont {Z.-L.}\ \bibnamefont
  {Deck-L{\'{e}}ger}},\ }\href@noop {} {\bibfield  {journal} {\bibinfo
  {journal} {IEEE Transactions on Antennas and Propagation}\ }\textbf {\bibinfo
  {volume} {68}},\ \bibinfo {pages} {1583} (\bibinfo {year}
  {2020})}\BibitemShut {NoStop}%
\bibitem [{\citenamefont {Bacot}\ \emph {et~al.}(2016)\citenamefont {Bacot},
  \citenamefont {Labousse}, \citenamefont {Eddi}, \citenamefont {Fink},\ and\
  \citenamefont {Fort}}]{Bacot2016}%
  \BibitemOpen
  \bibfield  {author} {\bibinfo {author} {\bibfnamefont {V.}~\bibnamefont
  {Bacot}}, \bibinfo {author} {\bibfnamefont {M.}~\bibnamefont {Labousse}},
  \bibinfo {author} {\bibfnamefont {A.}~\bibnamefont {Eddi}}, \bibinfo {author}
  {\bibfnamefont {M.}~\bibnamefont {Fink}}, \ and\ \bibinfo {author}
  {\bibfnamefont {E.}~\bibnamefont {Fort}},\ }\href@noop {} {\bibfield
  {journal} {\bibinfo  {journal} {Nature Physics}\ }\textbf {\bibinfo {volume}
  {12}},\ \bibinfo {pages} {972} (\bibinfo {year} {2016})}\BibitemShut
  {NoStop}%
\bibitem [{\citenamefont {Moussa}\ \emph {et~al.}(2023)\citenamefont {Moussa},
  \citenamefont {Xu}, \citenamefont {Yin}, \citenamefont {Galiffi},
  \citenamefont {Ra’di},\ and\ \citenamefont
  {Al{\`u}}}]{moussa2023observation}%
  \BibitemOpen
  \bibfield  {author} {\bibinfo {author} {\bibfnamefont {H.}~\bibnamefont
  {Moussa}}, \bibinfo {author} {\bibfnamefont {G.}~\bibnamefont {Xu}}, \bibinfo
  {author} {\bibfnamefont {S.}~\bibnamefont {Yin}}, \bibinfo {author}
  {\bibfnamefont {E.}~\bibnamefont {Galiffi}}, \bibinfo {author} {\bibfnamefont
  {Y.}~\bibnamefont {Ra’di}}, \ and\ \bibinfo {author} {\bibfnamefont
  {A.}~\bibnamefont {Al{\`u}}},\ }\href@noop {} {\bibfield  {journal} {\bibinfo
   {journal} {Nature Physics}\ }\textbf {\bibinfo {volume} {19}},\ \bibinfo
  {pages} {863} (\bibinfo {year} {2023})}\BibitemShut {NoStop}%
\bibitem [{\citenamefont {Akbarzadeh}\ \emph {et~al.}(2018)\citenamefont
  {Akbarzadeh}, \citenamefont {Chamanara},\ and\ \citenamefont
  {Caloz}}]{Akbarzadeh2018}%
  \BibitemOpen
  \bibfield  {author} {\bibinfo {author} {\bibfnamefont {A.}~\bibnamefont
  {Akbarzadeh}}, \bibinfo {author} {\bibfnamefont {N.}~\bibnamefont
  {Chamanara}}, \ and\ \bibinfo {author} {\bibfnamefont {C.}~\bibnamefont
  {Caloz}},\ }\href@noop {} {\bibfield  {journal} {\bibinfo  {journal} {Optics
  Letters}\ }\textbf {\bibinfo {volume} {43}},\ \bibinfo {pages} {3297}
  (\bibinfo {year} {2018})}\BibitemShut {NoStop}%
\bibitem [{\citenamefont {McCall}\ \emph {et~al.}(2010)\citenamefont {McCall},
  \citenamefont {Favaro}, \citenamefont {Kinsler},\ and\ \citenamefont
  {Boardman}}]{mccall2010spacetime}%
  \BibitemOpen
  \bibfield  {author} {\bibinfo {author} {\bibfnamefont {M.~W.}\ \bibnamefont
  {McCall}}, \bibinfo {author} {\bibfnamefont {A.}~\bibnamefont {Favaro}},
  \bibinfo {author} {\bibfnamefont {P.}~\bibnamefont {Kinsler}}, \ and\
  \bibinfo {author} {\bibfnamefont {A.}~\bibnamefont {Boardman}},\ }\href@noop
  {} {\bibfield  {journal} {\bibinfo  {journal} {Journal of Optics}\ }\textbf
  {\bibinfo {volume} {13}},\ \bibinfo {pages} {024003} (\bibinfo {year}
  {2010})}\BibitemShut {NoStop}%
\bibitem [{\citenamefont {Zeng}\ \emph {et~al.}(2017)\citenamefont {Zeng},
  \citenamefont {Xu}, \citenamefont {Wang}, \citenamefont {Zhang},
  \citenamefont {Zhao}, \citenamefont {Zeng},\ and\ \citenamefont
  {Song}}]{zeng2017photonic}%
  \BibitemOpen
  \bibfield  {author} {\bibinfo {author} {\bibfnamefont {L.}~\bibnamefont
  {Zeng}}, \bibinfo {author} {\bibfnamefont {J.}~\bibnamefont {Xu}}, \bibinfo
  {author} {\bibfnamefont {C.}~\bibnamefont {Wang}}, \bibinfo {author}
  {\bibfnamefont {J.}~\bibnamefont {Zhang}}, \bibinfo {author} {\bibfnamefont
  {Y.}~\bibnamefont {Zhao}}, \bibinfo {author} {\bibfnamefont {J.}~\bibnamefont
  {Zeng}}, \ and\ \bibinfo {author} {\bibfnamefont {R.}~\bibnamefont {Song}},\
  }\href@noop {} {\bibfield  {journal} {\bibinfo  {journal} {Scientific
  Reports}\ }\textbf {\bibinfo {volume} {7}},\ \bibinfo {pages} {1} (\bibinfo
  {year} {2017})}\BibitemShut {NoStop}%
\bibitem [{\citenamefont {Fleury}\ \emph {et~al.}(2014)\citenamefont {Fleury},
  \citenamefont {Sounas},\ and\ \citenamefont {Alu}}]{fleury2014negative}%
  \BibitemOpen
  \bibfield  {author} {\bibinfo {author} {\bibfnamefont {R.}~\bibnamefont
  {Fleury}}, \bibinfo {author} {\bibfnamefont {D.~L.}\ \bibnamefont {Sounas}},
  \ and\ \bibinfo {author} {\bibfnamefont {A.}~\bibnamefont {Alu}},\
  }\href@noop {} {\bibfield  {journal} {\bibinfo  {journal} {Physical Review
  Letters}\ }\textbf {\bibinfo {volume} {113}},\ \bibinfo {pages} {023903}
  (\bibinfo {year} {2014})}\BibitemShut {NoStop}%
\bibitem [{\citenamefont {Huidobro}\ \emph {et~al.}(2019)\citenamefont
  {Huidobro}, \citenamefont {Galiffi}, \citenamefont {Guenneau}, \citenamefont
  {Craster},\ and\ \citenamefont {Pendry}}]{huidobro2019fresnel}%
  \BibitemOpen
  \bibfield  {author} {\bibinfo {author} {\bibfnamefont {P.~A.}\ \bibnamefont
  {Huidobro}}, \bibinfo {author} {\bibfnamefont {E.}~\bibnamefont {Galiffi}},
  \bibinfo {author} {\bibfnamefont {S.}~\bibnamefont {Guenneau}}, \bibinfo
  {author} {\bibfnamefont {R.~V.}\ \bibnamefont {Craster}}, \ and\ \bibinfo
  {author} {\bibfnamefont {J.~B.}\ \bibnamefont {Pendry}},\ }\href@noop {}
  {\bibfield  {journal} {\bibinfo  {journal} {Proceedings of the National
  Academy of Sciences}\ }\textbf {\bibinfo {volume} {116}},\ \bibinfo {pages}
  {24943} (\bibinfo {year} {2019})}\BibitemShut {NoStop}%
\bibitem [{\citenamefont {Trainiti}\ \emph {et~al.}(2019)\citenamefont
  {Trainiti}, \citenamefont {Xia}, \citenamefont {Marconi}, \citenamefont
  {Cazzulani}, \citenamefont {Erturk},\ and\ \citenamefont
  {Ruzzene}}]{trainiti2019time}%
  \BibitemOpen
  \bibfield  {author} {\bibinfo {author} {\bibfnamefont {G.}~\bibnamefont
  {Trainiti}}, \bibinfo {author} {\bibfnamefont {Y.}~\bibnamefont {Xia}},
  \bibinfo {author} {\bibfnamefont {J.}~\bibnamefont {Marconi}}, \bibinfo
  {author} {\bibfnamefont {G.}~\bibnamefont {Cazzulani}}, \bibinfo {author}
  {\bibfnamefont {A.}~\bibnamefont {Erturk}}, \ and\ \bibinfo {author}
  {\bibfnamefont {M.}~\bibnamefont {Ruzzene}},\ }\href@noop {} {\bibfield
  {journal} {\bibinfo  {journal} {Physical Review Letters}\ }\textbf {\bibinfo
  {volume} {122}},\ \bibinfo {pages} {124301} (\bibinfo {year}
  {2019})}\BibitemShut {NoStop}%
\bibitem [{\citenamefont {Pacheco-Pe{\~n}a}\ and\ \citenamefont
  {Engheta}(2020)}]{pacheco2020temporal}%
  \BibitemOpen
  \bibfield  {author} {\bibinfo {author} {\bibfnamefont {V.}~\bibnamefont
  {Pacheco-Pe{\~n}a}}\ and\ \bibinfo {author} {\bibfnamefont {N.}~\bibnamefont
  {Engheta}},\ }\href@noop {} {\bibfield  {journal} {\bibinfo  {journal}
  {Light: Science \& Applications}\ }\textbf {\bibinfo {volume} {9}},\ \bibinfo
  {pages} {129} (\bibinfo {year} {2020})}\BibitemShut {NoStop}%
\bibitem [{\citenamefont {Kort-Kamp}\ \emph {et~al.}(2021)\citenamefont
  {Kort-Kamp}, \citenamefont {Azad},\ and\ \citenamefont
  {Dalvit}}]{kort2021space}%
  \BibitemOpen
  \bibfield  {author} {\bibinfo {author} {\bibfnamefont {W.~J.}\ \bibnamefont
  {Kort-Kamp}}, \bibinfo {author} {\bibfnamefont {A.~K.}\ \bibnamefont {Azad}},
  \ and\ \bibinfo {author} {\bibfnamefont {D.~A.}\ \bibnamefont {Dalvit}},\
  }\href@noop {} {\bibfield  {journal} {\bibinfo  {journal} {Physical Review
  Letters}\ }\textbf {\bibinfo {volume} {127}},\ \bibinfo {pages} {043603}
  (\bibinfo {year} {2021})}\BibitemShut {NoStop}%
\bibitem [{\citenamefont {Mora}\ \emph {et~al.}(2022)\citenamefont {Mora},
  \citenamefont {Chekroun}, \citenamefont {Raetz},\ and\ \citenamefont
  {Tournat}}]{mora2022nonlinear}%
  \BibitemOpen
  \bibfield  {author} {\bibinfo {author} {\bibfnamefont {P.}~\bibnamefont
  {Mora}}, \bibinfo {author} {\bibfnamefont {M.}~\bibnamefont {Chekroun}},
  \bibinfo {author} {\bibfnamefont {S.}~\bibnamefont {Raetz}}, \ and\ \bibinfo
  {author} {\bibfnamefont {V.}~\bibnamefont {Tournat}},\ }\href@noop {}
  {\bibfield  {journal} {\bibinfo  {journal} {Ultrasonics}\ }\textbf {\bibinfo
  {volume} {119}},\ \bibinfo {pages} {106589} (\bibinfo {year}
  {2022})}\BibitemShut {NoStop}%
\bibitem [{\citenamefont {Kiefer}\ \emph {et~al.}(2023)\citenamefont {Kiefer},
  \citenamefont {Plestenjak}, \citenamefont {Gravenkamp},\ and\ \citenamefont
  {Prada}}]{kiefer2023computing}%
  \BibitemOpen
  \bibfield  {author} {\bibinfo {author} {\bibfnamefont {D.~A.}\ \bibnamefont
  {Kiefer}}, \bibinfo {author} {\bibfnamefont {B.}~\bibnamefont {Plestenjak}},
  \bibinfo {author} {\bibfnamefont {H.}~\bibnamefont {Gravenkamp}}, \ and\
  \bibinfo {author} {\bibfnamefont {C.}~\bibnamefont {Prada}},\ }\href@noop {}
  {\bibfield  {journal} {\bibinfo  {journal} {The Journal of the Acoustical
  Society of America}\ }\textbf {\bibinfo {volume} {153}},\ \bibinfo {pages}
  {1386} (\bibinfo {year} {2023})}\BibitemShut {NoStop}%
\bibitem [{\citenamefont {Samak}\ and\ \citenamefont
  {Bilal}(2024)}]{samak2024evidence}%
  \BibitemOpen
  \bibfield  {author} {\bibinfo {author} {\bibfnamefont {M.~M.}\ \bibnamefont
  {Samak}}\ and\ \bibinfo {author} {\bibfnamefont {O.~R.}\ \bibnamefont
  {Bilal}},\ }\href@noop {} {\bibfield  {journal} {\bibinfo  {journal} {APL
  Materials}\ }\textbf {\bibinfo {volume} {12}} (\bibinfo {year}
  {2024})}\BibitemShut {NoStop}%
\bibitem [{\citenamefont {Liu}\ \emph {et~al.}(2001)\citenamefont {Liu},
  \citenamefont {Dutton}, \citenamefont {Behroozi},\ and\ \citenamefont
  {Hau}}]{liu2001observation}%
  \BibitemOpen
  \bibfield  {author} {\bibinfo {author} {\bibfnamefont {C.}~\bibnamefont
  {Liu}}, \bibinfo {author} {\bibfnamefont {Z.}~\bibnamefont {Dutton}},
  \bibinfo {author} {\bibfnamefont {C.~H.}\ \bibnamefont {Behroozi}}, \ and\
  \bibinfo {author} {\bibfnamefont {L.~V.}\ \bibnamefont {Hau}},\ }\href@noop
  {} {\bibfield  {journal} {\bibinfo  {journal} {Nature}\ }\textbf {\bibinfo
  {volume} {409}},\ \bibinfo {pages} {490} (\bibinfo {year}
  {2001})}\BibitemShut {NoStop}%
\bibitem [{\citenamefont {Figotin}\ and\ \citenamefont
  {Vitebskiy}(2003)}]{figotin2003oblique}%
  \BibitemOpen
  \bibfield  {author} {\bibinfo {author} {\bibfnamefont {A.}~\bibnamefont
  {Figotin}}\ and\ \bibinfo {author} {\bibfnamefont {I.}~\bibnamefont
  {Vitebskiy}},\ }\href@noop {} {\bibfield  {journal} {\bibinfo  {journal}
  {Physical Review E}\ }\textbf {\bibinfo {volume} {68}},\ \bibinfo {pages}
  {036609} (\bibinfo {year} {2003})}\BibitemShut {NoStop}%
\bibitem [{\citenamefont {Baba}(2008)}]{baba2008slow}%
  \BibitemOpen
  \bibfield  {author} {\bibinfo {author} {\bibfnamefont {T.}~\bibnamefont
  {Baba}},\ }\href@noop {} {\bibfield  {journal} {\bibinfo  {journal} {Nature
  Photonics}\ }\textbf {\bibinfo {volume} {2}},\ \bibinfo {pages} {465}
  (\bibinfo {year} {2008})}\BibitemShut {NoStop}%
\bibitem [{\citenamefont {Khurgin}(2010)}]{khurgin2010slow}%
  \BibitemOpen
  \bibfield  {author} {\bibinfo {author} {\bibfnamefont {J.~B.}\ \bibnamefont
  {Khurgin}},\ }\href@noop {} {\bibfield  {journal} {\bibinfo  {journal}
  {Advances in Optics and Photonics}\ }\textbf {\bibinfo {volume} {2}},\
  \bibinfo {pages} {287} (\bibinfo {year} {2010})}\BibitemShut {NoStop}%
\bibitem [{\citenamefont {Figotin}\ and\ \citenamefont
  {Vitebskiy}(2011)}]{figotin2011slow}%
  \BibitemOpen
  \bibfield  {author} {\bibinfo {author} {\bibfnamefont {A.}~\bibnamefont
  {Figotin}}\ and\ \bibinfo {author} {\bibfnamefont {I.}~\bibnamefont
  {Vitebskiy}},\ }\href@noop {} {\bibfield  {journal} {\bibinfo  {journal}
  {Laser \& Photonics Reviews}\ }\textbf {\bibinfo {volume} {5}},\ \bibinfo
  {pages} {201} (\bibinfo {year} {2011})}\BibitemShut {NoStop}%
\bibitem [{\citenamefont {Goldzak}\ \emph {et~al.}(2018)\citenamefont
  {Goldzak}, \citenamefont {Mailybaev},\ and\ \citenamefont
  {Moiseyev}}]{goldzak2018light}%
  \BibitemOpen
  \bibfield  {author} {\bibinfo {author} {\bibfnamefont {T.}~\bibnamefont
  {Goldzak}}, \bibinfo {author} {\bibfnamefont {A.~A.}\ \bibnamefont
  {Mailybaev}}, \ and\ \bibinfo {author} {\bibfnamefont {N.}~\bibnamefont
  {Moiseyev}},\ }\href@noop {} {\bibfield  {journal} {\bibinfo  {journal}
  {Physical Review Letters}\ }\textbf {\bibinfo {volume} {120}},\ \bibinfo
  {pages} {013901} (\bibinfo {year} {2018})}\BibitemShut {NoStop}%
\bibitem [{\citenamefont {Yamilov}\ \emph {et~al.}(2023)\citenamefont
  {Yamilov}, \citenamefont {Skipetrov}, \citenamefont {Hughes}, \citenamefont
  {Minkov}, \citenamefont {Yu},\ and\ \citenamefont
  {Cao}}]{yamilov2023anderson}%
  \BibitemOpen
  \bibfield  {author} {\bibinfo {author} {\bibfnamefont {A.}~\bibnamefont
  {Yamilov}}, \bibinfo {author} {\bibfnamefont {S.~E.}\ \bibnamefont
  {Skipetrov}}, \bibinfo {author} {\bibfnamefont {T.~W.}\ \bibnamefont
  {Hughes}}, \bibinfo {author} {\bibfnamefont {M.}~\bibnamefont {Minkov}},
  \bibinfo {author} {\bibfnamefont {Z.}~\bibnamefont {Yu}}, \ and\ \bibinfo
  {author} {\bibfnamefont {H.}~\bibnamefont {Cao}},\ }\href@noop {} {\bibfield
  {journal} {\bibinfo  {journal} {Nature Physics}\ }\textbf {\bibinfo {volume}
  {19}},\ \bibinfo {pages} {1308} (\bibinfo {year} {2023})}\BibitemShut
  {NoStop}%
\bibitem [{\citenamefont {Figotin}\ and\ \citenamefont
  {Vitebsky}(2001)}]{figotin2001nonreciprocal}%
  \BibitemOpen
  \bibfield  {author} {\bibinfo {author} {\bibfnamefont {A.}~\bibnamefont
  {Figotin}}\ and\ \bibinfo {author} {\bibfnamefont {I.}~\bibnamefont
  {Vitebsky}},\ }\href@noop {} {\bibfield  {journal} {\bibinfo  {journal}
  {Physical Review E}\ }\textbf {\bibinfo {volume} {63}},\ \bibinfo {pages}
  {066609} (\bibinfo {year} {2001})}\BibitemShut {NoStop}%
\bibitem [{\citenamefont {Figotin}\ and\ \citenamefont
  {Vitebskiy}(2006)}]{figotin2006slow}%
  \BibitemOpen
  \bibfield  {author} {\bibinfo {author} {\bibfnamefont {A.}~\bibnamefont
  {Figotin}}\ and\ \bibinfo {author} {\bibfnamefont {I.}~\bibnamefont
  {Vitebskiy}},\ }\href@noop {} {\bibfield  {journal} {\bibinfo  {journal}
  {Waves in Random and Complex Media}\ }\textbf {\bibinfo {volume} {16}},\
  \bibinfo {pages} {293} (\bibinfo {year} {2006})}\BibitemShut {NoStop}%
\bibitem [{\citenamefont {Ballato}\ \emph {et~al.}(2005)\citenamefont
  {Ballato}, \citenamefont {Ballato}, \citenamefont {Figotin},\ and\
  \citenamefont {Vitebskiy}}]{ballato2005frozen}%
  \BibitemOpen
  \bibfield  {author} {\bibinfo {author} {\bibfnamefont {J.}~\bibnamefont
  {Ballato}}, \bibinfo {author} {\bibfnamefont {A.}~\bibnamefont {Ballato}},
  \bibinfo {author} {\bibfnamefont {A.}~\bibnamefont {Figotin}}, \ and\
  \bibinfo {author} {\bibfnamefont {I.}~\bibnamefont {Vitebskiy}},\ }\href@noop
  {} {\bibfield  {journal} {\bibinfo  {journal} {Physical Review E}\ }\textbf
  {\bibinfo {volume} {71}},\ \bibinfo {pages} {036612} (\bibinfo {year}
  {2005})}\BibitemShut {NoStop}%
\bibitem [{\citenamefont {Chen}\ \emph {et~al.}(2021)\citenamefont {Chen},
  \citenamefont {Kadic},\ and\ \citenamefont {Wegener}}]{chen2021roton}%
  \BibitemOpen
  \bibfield  {author} {\bibinfo {author} {\bibfnamefont {Y.}~\bibnamefont
  {Chen}}, \bibinfo {author} {\bibfnamefont {M.}~\bibnamefont {Kadic}}, \ and\
  \bibinfo {author} {\bibfnamefont {M.}~\bibnamefont {Wegener}},\ }\href@noop
  {} {\bibfield  {journal} {\bibinfo  {journal} {Nature Communications}\
  }\textbf {\bibinfo {volume} {12}},\ \bibinfo {pages} {3278} (\bibinfo {year}
  {2021})}\BibitemShut {NoStop}%
\bibitem [{\citenamefont {Wang}\ \emph {et~al.}(2022)\citenamefont {Wang},
  \citenamefont {Chen}, \citenamefont {Kadic}, \citenamefont {Wang},\ and\
  \citenamefont {Wegener}}]{wang2022nonlocal}%
  \BibitemOpen
  \bibfield  {author} {\bibinfo {author} {\bibfnamefont {K.}~\bibnamefont
  {Wang}}, \bibinfo {author} {\bibfnamefont {Y.}~\bibnamefont {Chen}}, \bibinfo
  {author} {\bibfnamefont {M.}~\bibnamefont {Kadic}}, \bibinfo {author}
  {\bibfnamefont {C.}~\bibnamefont {Wang}}, \ and\ \bibinfo {author}
  {\bibfnamefont {M.}~\bibnamefont {Wegener}},\ }\href@noop {} {\bibfield
  {journal} {\bibinfo  {journal} {Communications Materials}\ }\textbf {\bibinfo
  {volume} {3}},\ \bibinfo {pages} {35} (\bibinfo {year} {2022})}\BibitemShut
  {NoStop}%
\bibitem [{\citenamefont {Kazemi}\ \emph {et~al.}(2023)\citenamefont {Kazemi},
  \citenamefont {Deshmukh}, \citenamefont {Chen}, \citenamefont {Liu},
  \citenamefont {Deng}, \citenamefont {Fu},\ and\ \citenamefont
  {Wang}}]{kazemi2023drawing}%
  \BibitemOpen
  \bibfield  {author} {\bibinfo {author} {\bibfnamefont {A.}~\bibnamefont
  {Kazemi}}, \bibinfo {author} {\bibfnamefont {K.~J.}\ \bibnamefont
  {Deshmukh}}, \bibinfo {author} {\bibfnamefont {F.}~\bibnamefont {Chen}},
  \bibinfo {author} {\bibfnamefont {Y.}~\bibnamefont {Liu}}, \bibinfo {author}
  {\bibfnamefont {B.}~\bibnamefont {Deng}}, \bibinfo {author} {\bibfnamefont
  {H.~C.}\ \bibnamefont {Fu}}, \ and\ \bibinfo {author} {\bibfnamefont
  {P.}~\bibnamefont {Wang}},\ }\href@noop {} {\bibfield  {journal} {\bibinfo
  {journal} {Physical Review Letters}\ }\textbf {\bibinfo {volume} {131}},\
  \bibinfo {pages} {176101} (\bibinfo {year} {2023})}\BibitemShut {NoStop}%
\bibitem [{\citenamefont {Bloch}(1929)}]{bloch1929quantenmechanik}%
  \BibitemOpen
  \bibfield  {author} {\bibinfo {author} {\bibfnamefont {F.}~\bibnamefont
  {Bloch}},\ }\href@noop {} {\bibfield  {journal} {\bibinfo  {journal}
  {Zeitschrift f{\"u}r physik}\ }\textbf {\bibinfo {volume} {52}},\ \bibinfo
  {pages} {555} (\bibinfo {year} {1929})}\BibitemShut {NoStop}%
\bibitem [{\citenamefont {Remoissenet}(2013)}]{remoissenet2013waves}%
  \BibitemOpen
  \bibfield  {author} {\bibinfo {author} {\bibfnamefont {M.}~\bibnamefont
  {Remoissenet}},\ }\href@noop {} {\emph {\bibinfo {title} {Waves called
  solitons: concepts and experiments}}}\ (\bibinfo  {publisher} {Springer
  Science \& Business Media},\ \bibinfo {year} {2013})\BibitemShut {NoStop}%
\bibitem [{SI()}]{SI}%
  \BibitemOpen
  \href@noop {} {\enquote {\bibinfo {title} {See supplemental information at
  url for additional results, and discussions. the supplemental information
  cites refs. [cite]},}\ }\BibitemShut {NoStop}%
\bibitem [{\citenamefont {Deshmukh}\ and\ \citenamefont
  {Milton}(2022)}]{deshmukh2022energy}%
  \BibitemOpen
  \bibfield  {author} {\bibinfo {author} {\bibfnamefont {K.~J.}\ \bibnamefont
  {Deshmukh}}\ and\ \bibinfo {author} {\bibfnamefont {G.~W.}\ \bibnamefont
  {Milton}},\ }\href@noop {} {\bibfield  {journal} {\bibinfo  {journal}
  {Applied Physics Letters}\ }\textbf {\bibinfo {volume} {121}} (\bibinfo
  {year} {2022})}\BibitemShut {NoStop}%
\end{thebibliography}%


\begin{thebibliography}{5}%
\makeatletter
\providecommand \@ifxundefined [1]{%
 \@ifx{#1\undefined}
}%
\providecommand \@ifnum [1]{%
 \ifnum #1\expandafter \@firstoftwo
 \else \expandafter \@secondoftwo
 \fi
}%
\providecommand \@ifx [1]{%
 \ifx #1\expandafter \@firstoftwo
 \else \expandafter \@secondoftwo
 \fi
}%
\providecommand \natexlab [1]{#1}%
\providecommand \enquote  [1]{``#1''}%
\providecommand \bibnamefont  [1]{#1}%
\providecommand \bibfnamefont [1]{#1}%
\providecommand \citenamefont [1]{#1}%
\providecommand \href@noop [0]{\@secondoftwo}%
\providecommand \href [0]{\begingroup \@sanitize@url \@href}%
\providecommand \@href[1]{\@@startlink{#1}\@@href}%
\providecommand \@@href[1]{\endgroup#1\@@endlink}%
\providecommand \@sanitize@url [0]{\catcode `\\12\catcode `\$12\catcode
  `\&12\catcode `\#12\catcode `\^12\catcode `\_12\catcode `\%12\relax}%
\providecommand \@@startlink[1]{}%
\providecommand \@@endlink[0]{}%
\providecommand \url  [0]{\begingroup\@sanitize@url \@url }%
\providecommand \@url [1]{\endgroup\@href {#1}{\urlprefix }}%
\providecommand \urlprefix  [0]{URL }%
\providecommand \Eprint [0]{\href }%
\providecommand \doibase [0]{http://dx.doi.org/}%
\providecommand \selectlanguage [0]{\@gobble}%
\providecommand \bibinfo  [0]{\@secondoftwo}%
\providecommand \bibfield  [0]{\@secondoftwo}%
\providecommand \translation [1]{[#1]}%
\providecommand \BibitemOpen [0]{}%
\providecommand \bibitemStop [0]{}%
\providecommand \bibitemNoStop [0]{.\EOS\space}%
\providecommand \EOS [0]{\spacefactor3000\relax}%
\providecommand \BibitemShut  [1]{\csname bibitem#1\endcsname}%
\let\auto@bib@innerbib\@empty
%</preamble>
\bibitem [{\citenamefont {Deshmukh}\ and\ \citenamefont
  {Milton}(2022)}]{deshmukh2022energy}%
  \BibitemOpen
  \bibfield  {author} {\bibinfo {author} {\bibfnamefont {K.~J.}\ \bibnamefont
  {Deshmukh}}\ and\ \bibinfo {author} {\bibfnamefont {G.~W.}\ \bibnamefont
  {Milton}},\ }\bibfield  {title} {\enquote {\bibinfo {title} {An energy
  conserving mechanism for temporal metasurfaces},}\ }\href@noop {} {\bibfield
  {journal} {\bibinfo  {journal} {Applied Physics Letters}\ }\textbf {\bibinfo
  {volume} {121}} (\bibinfo {year} {2022})}\BibitemShut {NoStop}%
\bibitem [{\citenamefont {Lurie}(2007)}]{lurie2007introduction}%
  \BibitemOpen
  \bibfield  {author} {\bibinfo {author} {\bibfnamefont {K.~A.}\ \bibnamefont
  {Lurie}},\ }\href@noop {} {\emph {\bibinfo {title} {An introduction to the
  mathematical theory of dynamic materials}}},\ Vol.~\bibinfo {volume} {15}\
  (\bibinfo  {publisher} {Springer},\ \bibinfo {year} {2007})\BibitemShut
  {NoStop}%
\bibitem [{\citenamefont {Lurie}\ and\ \citenamefont
  {Yakovlev}(2016)}]{Lurie2016}%
  \BibitemOpen
  \bibfield  {author} {\bibinfo {author} {\bibfnamefont {K.~A.}\ \bibnamefont
  {Lurie}}\ and\ \bibinfo {author} {\bibfnamefont {V.~V.}\ \bibnamefont
  {Yakovlev}},\ }\bibfield  {title} {\enquote {\bibinfo {title} {{Energy
  accumulation in waves propagating in space and time-varying transmission
  lines}},}\ }\href@noop {} {\bibfield  {journal} {\bibinfo  {journal} {IEEE
  Antennas and Wireless Propagation Letters}\ }\textbf {\bibinfo {volume}
  {15}},\ \bibinfo {pages} {1681--1684} (\bibinfo {year} {2016})}\BibitemShut
  {NoStop}%
\bibitem [{\citenamefont {Milton}\ and\ \citenamefont
  {Mattei}(2017)}]{milton2017field}%
  \BibitemOpen
  \bibfield  {author} {\bibinfo {author} {\bibfnamefont {G.~W.}\ \bibnamefont
  {Milton}}\ and\ \bibinfo {author} {\bibfnamefont {O.}~\bibnamefont
  {Mattei}},\ }\bibfield  {title} {\enquote {\bibinfo {title} {Field patterns:
  a new mathematical object},}\ }\href@noop {} {\bibfield  {journal} {\bibinfo
  {journal} {Proceedings of the Royal Society A: Mathematical, Physical and
  Engineering Sciences}\ }\textbf {\bibinfo {volume} {473}},\ \bibinfo {pages}
  {20160819} (\bibinfo {year} {2017})}\BibitemShut {NoStop}%
\bibitem [{\citenamefont {Mattei}\ and\ \citenamefont
  {Milton}(2018)}]{mattei2018field}%
  \BibitemOpen
  \bibfield  {author} {\bibinfo {author} {\bibfnamefont {O.}~\bibnamefont
  {Mattei}}\ and\ \bibinfo {author} {\bibfnamefont {G.~W.}\ \bibnamefont
  {Milton}},\ }\bibfield  {title} {\enquote {\bibinfo {title} {Field patterns:
  a new type of wave with infinitely degenerate band structure},}\ }\href@noop
  {} {\bibfield  {journal} {\bibinfo  {journal} {Europhysics Letters}\ }\textbf
  {\bibinfo {volume} {120}},\ \bibinfo {pages} {54003} (\bibinfo {year}
  {2018})}\BibitemShut {NoStop}%
\end{thebibliography}%
%TC:endignore

\end{document}

\preprint{Supplementary Information}
\title[Wave-freezing in Temporal Metasurfaces]{Supplementary Information: Wave-Freezing and other Phenomena in Temporal Metasurfaces driven by Nonlocal Interactions.} 
% Force line breaks with \\
\affiliation{ 
Department of Mathematics, University of Utah, Salt Lake City, Utah 84112, USA}%
\author{Kshiteej J. Deshmukh\orcidlink{0000-0002-6825-4280}\thanks{*}}
\email[Corresponding author: ]{kjdeshmu@math.utah.edu}


%

% \homepage{http://www.Second.institution.edu/~Charlie.Author.}
%\affiliation{%
%Second institution and/or address%\\This line break forced% with \\
%}%
\date{\today}% It is always \today, today,
{
\let\clearpage\relax
\maketitle
}

%%%%%%%%%%%%%%%%%%%%%%%%%
%%%%%%%%%%%%%%%%%%%%%%%%%
\vspace{-1cm}
    This document provides the supplementary information for the main manuscript. 
%%%%%%%%%%%%%%%%%%%%%%%%%
%%%%%%%%%%%%%%%%%%%%%%%%%
%%%%%%%%%%%%%%%%%%%%%%%%%
%%%%%%%%%%%%%%%%%%%%%%%%%
%%%%%%%%%%%%%%%%%%%%%%%%%
%%%%%%%%%%%%%%%%%%%%%%%%%
%%%%%%%%%%%%%%%%%%%%
%%%%%%%%%%%%%%%%%%%%
%%%%%%%%%%%%%%%%%%%%
%%%%%%%%%%%%%%%%%%%%
\section{One-dimensional temporal metasurfaces}
\subsection{Energy exchange in one-dimensional temporal metasurfaces for wave-freezing}
%%%
%%% WAVE-FREEZING energy exchange
\begin{figure}[bp!]
        \centering
        \includegraphics[width=\linewidth]{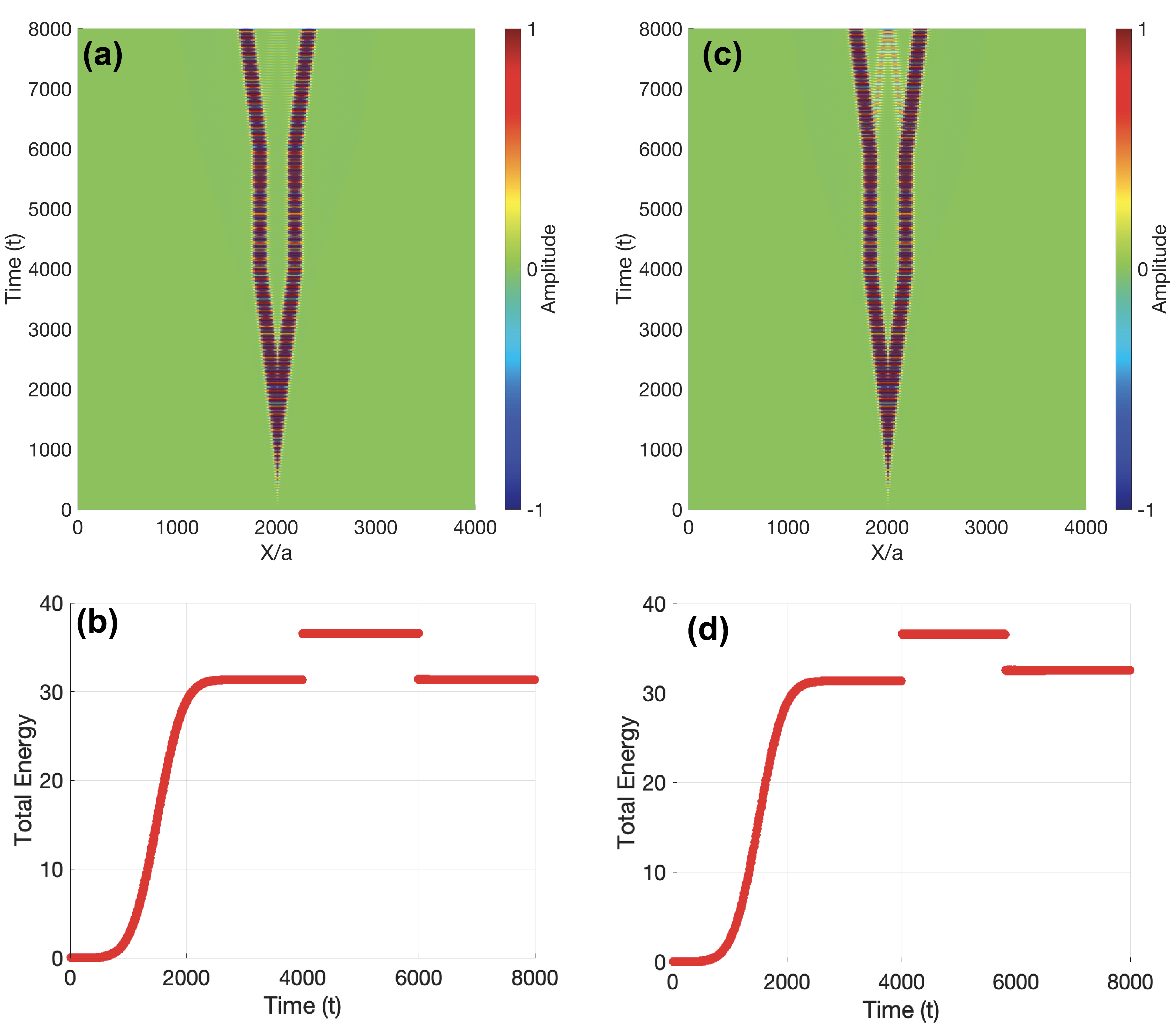}
\caption{Two cases of wave-freezing in 1-D are shown, where the only difference between the two cases is the instant of time at which the second time-interface is applied.
(a),(c) $X-t$ plots of the wave displacement amplitudes for the two cases.
(b), (d) Total energy of the system vs. time plots for the two cases in (a) and (b), respectively.
}
\label{fig:energy_wave_freezing}
\end{figure}
Here, we present the energy exchange taking place in the 1-D temporal metasurfaces used for the \textit{wave-freezing} application.
Figure \ref{fig:energy_wave_freezing} shows two cases of \textit{wave-freezing} in 1-D.
It is well-known that time-modulation of material properties is typically associated with energy being added to the system or energy being removed from the system (except, in some energy conserving temporal metasurfaces\cite{deshmukh2022energy}).
In both cases, we consider a spring-mass chain consisting of $4000$ unit cells, with a forcing $f(t)$ applied to the center mass of the chain.
Initially, the spring-mass chain has local (nearest-neighbor) interactions only, with all masses having the same mass value $m=1$, and the spring stiffness values given by $k_1=0.01$.
The forcing function has a Gaussian envelop and is given by the expression, 
\begin{equation}\label{1D forcing}
    f(t) = A e^{-(t-1500)^2/\tau^2} \cos{(\omega_1 t)}.
\end{equation}
Here, $A=0.1$ is the forcing amplitude, $\omega_1 =0.1414$ is the frequency of the forcing, and $\tau=100/\omega_1$ is the parameter controlling the width of the Gaussian envelop.
This generates a wave propagating in the system with the frequency-wavevector value $(\omega, \kappa) = (0.1414, 0.5)$.

In both cases, the first time-interface is applied at $t=4000$, at which tailored third-nearest-neighbor interactions, modeled by nonlocal spring with stiffness $k_3=k_1/3$, are introduced in the system so that the resulting nonlocal system has $\parderivsec{\omega}{\kappa}\Big|_{\kappa=0.5}=0$, i.e., an inflection point, at the wavevector value of $\kappa=0.5$.
This causes the propagating wave to freeze in both cases, as both the systems are similar until this point, as shown in Figs.  \ref{fig:energy_wave_freezing}\red{(a)} and \ref{fig:energy_wave_freezing}\red{(c)}.
The two cases differ from each other in the instant of time at which we apply the $2^{nd}$ time-interface.
In the first case (Fig. \ref{fig:energy_wave_freezing}\red(a),(b)), the $2^{nd}$ time-interface is applied at $t=6000$, and in the second case (Fig. \ref{fig:energy_wave_freezing}\red(c),(d)) it is applied at a slightly different instant of time, $t=5980$.
At the $2^{nd}$ interface, the value of $k_3$ is reduced to $0$, and the spring-mass chain now has nearest-neighbor interactions only.
The \textit{frozen} wave starts propagating again by splitting into a forward propagating and a backward propagating wave.

Figures. \ref{fig:energy_wave_freezing}\red{(b)} and \ref{fig:energy_wave_freezing}\red{(d)} show the total energy of the system as it changes with respect to time for the two cases. 
We observe that initially the total energy of the spring-mass system is increasing due to the forcing applied to the center mass. 
The total energy then remains constant for a while after the forcing amplitude is reduced to $0$ as a result of the Gaussian envelop of the forcing function.
At the $1^{st}$ time-interface we observe a sudden increase in the total energy associated with the introduction of nonlocal springs with stiffness value $k_3$.
Let this increase in energy at the $1^{st}$ time-interface be denoted by $\Delta E_1$.
At the $2^{nd}$ time-interface, the total energy suddenly decreases by an amount $\Delta E_2$ as $k_3$ changes to $0$.
The amount of decrease in total energy is different ($\Delta E_1 \neq \Delta E_2$) in the two cases, and depends on the instant of time at which the $2^{nd}$ time-interface is applied. 
This is seen from the final total energies in Figs. \ref{fig:energy_wave_freezing}\red{(b)} and \red{(d)}.
As the nonlocal springs may be stretched or compressed by different amounts at different instants of time, it will result in different final total energies of the system. 
Interestingly, we observe that in the first case, when the final total energy of the system is very close to the total energy before the first time-interface, i.e., $\Delta E_1 \approx \Delta E_2$, the backward propagating wave has a negligible amplitude as compared to the second case, where the final total energy is slightly more than the total energy before the $1^{st}$ time-interface, i.e., $\Delta E_1 > \Delta E_2$,  and we observe a noticeable amplitude of the backward propagating wave.

This leads us to the interesting question: Can we eliminate backward propagating waves that appear after the $2^{nd}$ time-interface by appropriately choosing the instant of the $2^{nd}$ time-interface, and what is its relation to the total energy exchange happening in the system? 
This question needs further investigation to be answered concretely, and is a topic of further research.

%%%%%%%%
%%% Double-interface 1-D example from main paper
%%%%%%%%
\subsection{Time-reversed waves in one-dimensional temporal metasurface}
\begin{figure}[bp!]
        \centering
        \includegraphics[width=0.6\linewidth]{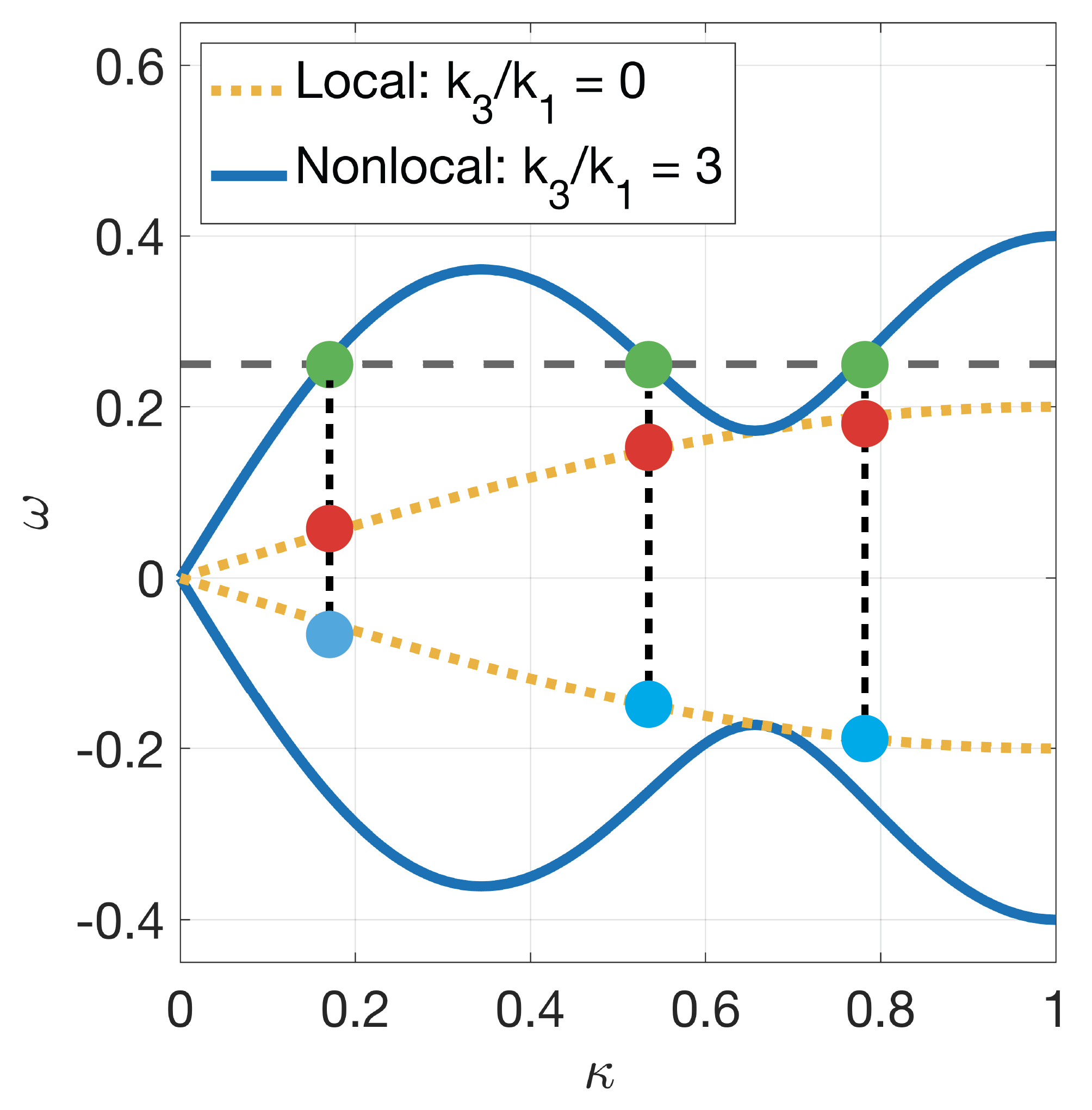}
\caption{Dispersion curves for the temporal metasurface of Fig. \textcolor{red}{2} in the main manuscript.
Before the time-interface, the spring-mass chain has nearest and third-nearest neighbor interactions with the dispersion relation shown by the solid blue curve.
The initial wave propagating in the nonlocal chain has three modes shown by the green markers.
At the first time-interface, the nonlocal interactions are reduced to $0$, and the spring-mass chain has local interactions only with the dispersion relation shown by the dotted yellow curve. 
The red and blue markers are the $(\omega, \kappa)$ values of the three modes in the local spring-mass chain.
}
\label{fig:1d_3modes}
\end{figure}
Figure \ref{fig:1d_3modes} shows the dispersion curves of the local and nonlocal spring-mass chains that form the temporal metasurface used for the 1-D example of Fig. \red{2} in the main manuscript showing image reconstruction from time-reversed waves.
Here, the center mass of the spring-mass chain with $18000$ unit cells is excited with a forcing that has a Gaussian envelop in time and is of the form given by Eqn. \eqref{1D forcing}, with $A=0.1$, $\omega_1 =0.25$, and $\tau=50/\omega_1$.

%%%%%%%%
%%%% Single-interface 1-D example for SI only
%%%%%%%%
\begin{figure}[bp!]
        \centering
        \includegraphics[width=\linewidth]{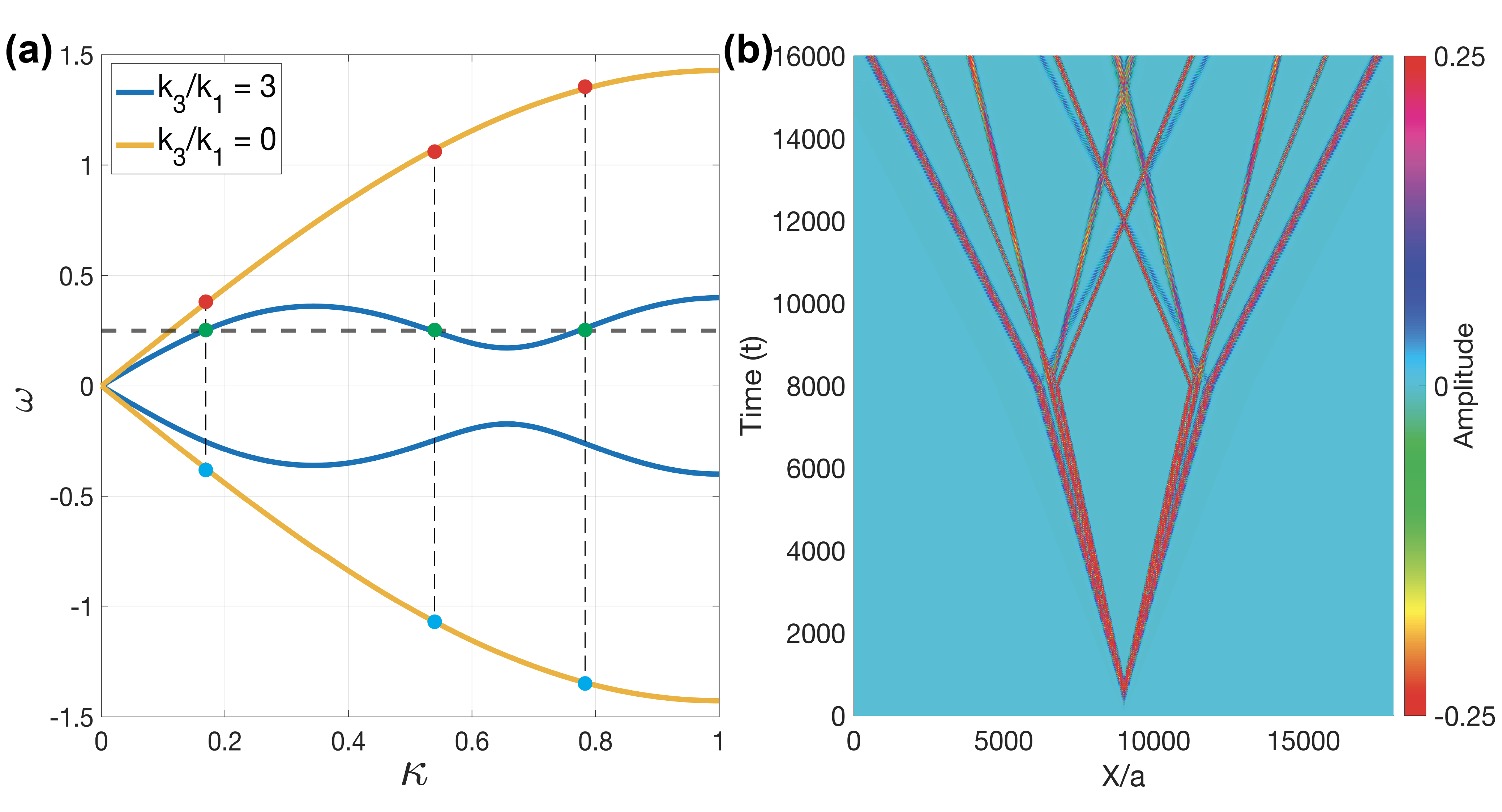}
\caption{Time-reversed waves creating multiple images of the source in a 1-D metasurface with a single time-interface: 
(a) Dispersion curves $\omega$ vs. $\kappa$ for the temporal metasurface which has nearest and third-nearest neighbor interactions (solid blue curve) before the time-interface, and only nearest-neighbor interactions (dotted yellow curve) after the time interface.
The green circle markers represent the 3 wave modes propagating at the same $\omega$ value, but different $\kappa$ values before the $1^{st}$ time-interface.
The red and blue markers represent the forward propagating and the backward propagating waves, respectively.
(b) $X-t$ plot of the displacement amplitude showing multiple reconstructed images. 
Colorbar represents displacement amplitude.
}
\label{fig:1d_3_images}
\end{figure}

We now demonstrate the case where a single time-interface is used to observe multiple reconstructed images from the time-reversed waves in a 1-D temporal metasurface. 
Consider a 1-d spring-mass chain that has nonlocal interactions to begin with, we choose, $N=3$, with $k_1 = 0.01, k_2 =0, k_3/k_1=3$. 
The dispersion relation of this nonlocal spring-mass chain is shown in Fig. \ref{fig:1d_3_images}\red{(a)} by the solid blue curve. 
The center mass of the spring-mass chain is subjected to a forcing of the form given in \eqref{1D forcing} with $A=0.1$, $\omega_1=0.25$, and $\tau=50/\omega_1$.
The forced oscillation produces 3 modes traveling to the right and 3 modes traveling to the left. 
The 3 right traveling modes are represented by the green circle markers in Fig. \ref{fig:1d_3_images}\red{(a)}, which show that they have the same $\omega$, but different $\kappa$ values.
At $t=9998$, we model the $1^{st}$ time interface, where the stiffness values of the nonlocal springs are reduced to $0$, i.e., $k_3=0$, and the stiffness of local springs is increased to $k_1=0.51$.
The $X-t$ plot of the displacement amplitude is  shown in Fig. \ref{fig:1d_3_images}\textcolor{red}{(b)}, where the colorbar represents the displacement amplitude.
When the propagating modes encounter the $1^{st}$ time interface, they are propagating in a material that has local (nearest-neighbor) interactions only (dispersion relation shown by yellow curve in Fig. \ref{fig:1d_3_images}\textcolor{red}{(a)}), and the 3 propagating modes now split  into modes which have different $\omega$ and $\kappa$ values (shown by red and blue markers in Fig. \ref{fig:1d_3_images}\textcolor{red}{(a)}).
The modes corresponding to the red markers form the forward propagating wave and the modes corresponding to the blue markers form the backward propagating wave.
The backward propagating wave, consisting of 3 modes with different $(\omega, \kappa)$ values, reconstructs to form three images of the initial wave-packet response.
Each of the modes has a distinct value of the pair $(\omega, \kappa)$, that results in each mode traveling with a different group-velocity as determined by the point on the dispersion curve.
The different group-velocities of  the time-reversed waves result in multiple images, three in this case (two of the modes have group-velocity values very close to each other and can be seen to form almost overlapping images).
Thus, using a single time-interface in the 1-D spring-mass chain, given a wave propagating at single frequency, we can obtain multiple reconstructed images  at different instants of time from waves with different frequencies.

%%%%%%%%%%%%%%%
%%%%%%%%%%%%%%%
%%%%%%%%%%%%%%%
%%%%%%%%%%%%%%%
%%%%%%%%%%%%%%%
%%%%%%%%%%%%%%%
\section{Two-dimensional temporal metasurfaces}
\begin{figure}[htp!]
        \centering
        \includegraphics[width=0.5\linewidth]{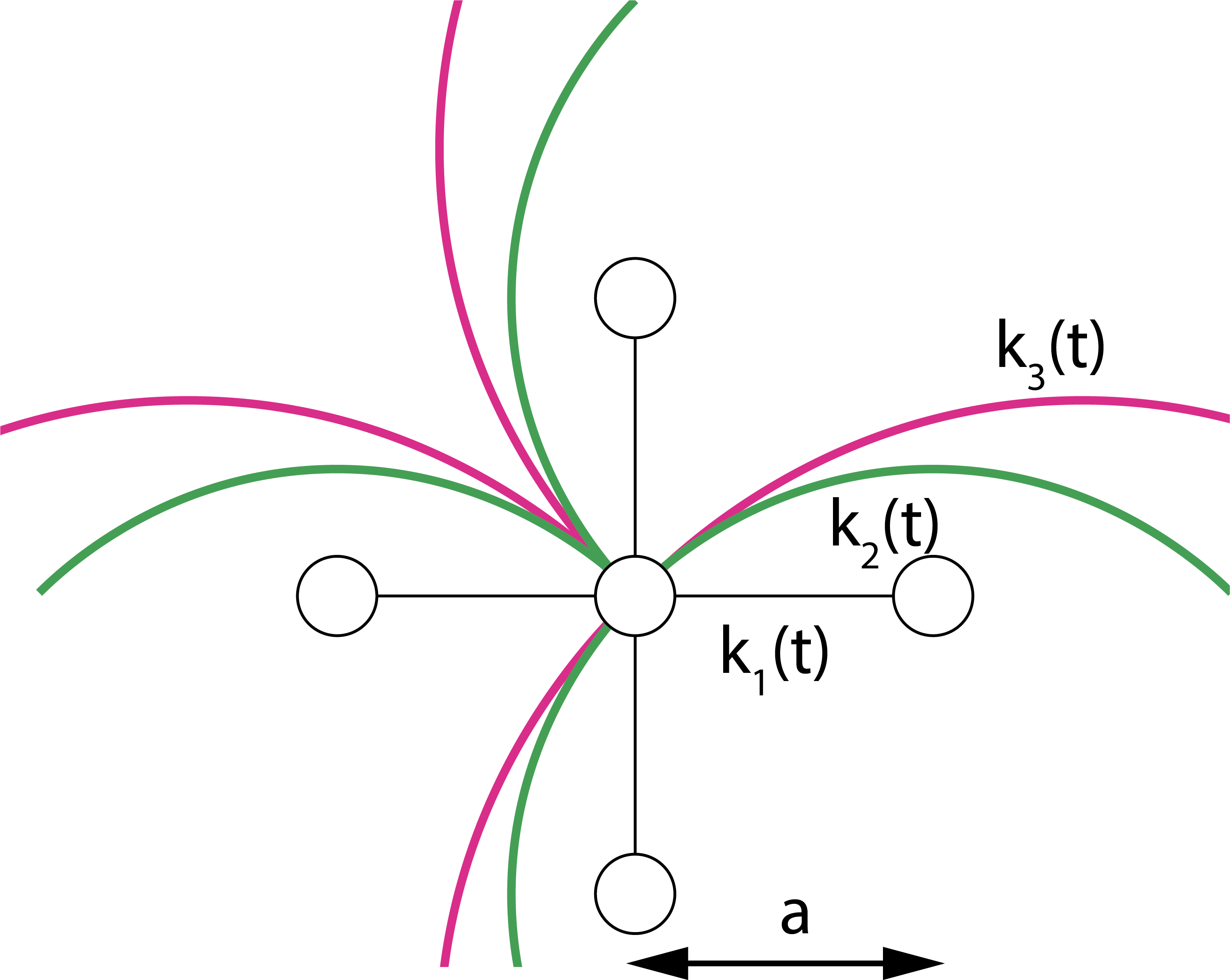}
\caption{Unit cell for the 2-D spring-mass system with nearest, second-nearest, and third-nearest neighbor interactions modeled by springs with stiffness values given by $k_1(t), k_2(t)$, and $k_3(t)$, respectively, in both, the $X$ and $Y$ directions.
The spring-mass system models the 2-D temporal metasurface of phononic crystals that have the strength of nonlocal interactions changing discontinuously with respect to time.
}
\label{fig:SI-Unit cell}
\end{figure}
Figure \ref{fig:SI-Unit cell} shows the unit cell for the two-dimensional spring-mass systems used in the 2-D numerical simulations presented in the main manuscript.
The 2-D unit cell is a square lattice with lattice constant $a=\pi$.
The schematic shows a mass interacting with the first, second, and third-nearest masses in both, the $X$, and $Y$ directions.
The examples shown in Fig. \red{3} and Fig. \red{4} of the main manuscript, consider the same number of nonlocal interactions in the $X$ and $Y$ directions, i.e., $N=M$.
The nonlocal interactions are modeled by springs with stiffness values given by $k_{x_1}(t)=k_{y_1}(t)\equiv k_1(t), k_{x_2}(t)=k_{y_2}(t) \equiv k_2(t)$, and $k_{x_3}(t)=k_{y_3}(t)\equiv k_3(t)$.

In general, the number of nonlocal interactions in the $X$ and $Y$ directions can be different, i.e, $N$ need not be equal to $M$.
The scalar displacements of the masses at the location $(x_i, y_j)$ is denoted by $u_{i,j}(t)$.
With all the masses having the same mass value $m$, the equation of motion for the displacement $u_{i,j}$ is given by, 
\begin{equation}\label{2d_DR}
    m \Ddot{u}_{i,j}= 
                        \sum_{p=1}^N 
                        k_{x_p} \left(u_{i+p,j}-2u_{i,j}+u_{i-p,j}\right)
                        +
                        \sum_{q=1}^M k_{y_q} 
                \left(u_{i,j+q}-2u_{i,j}+u_{i,j-q}\right),
\end{equation}

For the 2-D examples of \textit{wave-freezing} and anomalous temporal refraction shown in Figs. \red{3} and \red{4} of the main manuscript,  a finite difference code was implemented in MATLAB to solve initial value problems. 
A sufficiently large domain of $300\times300$ unit cells is considered with fixed-end boundary conditions, so that reflection of waves from the boundary does not affect the observations to be made from the numerical experiments.

\begin{figure}[tp!]
        \centering
        \includegraphics[width=0.8\linewidth]{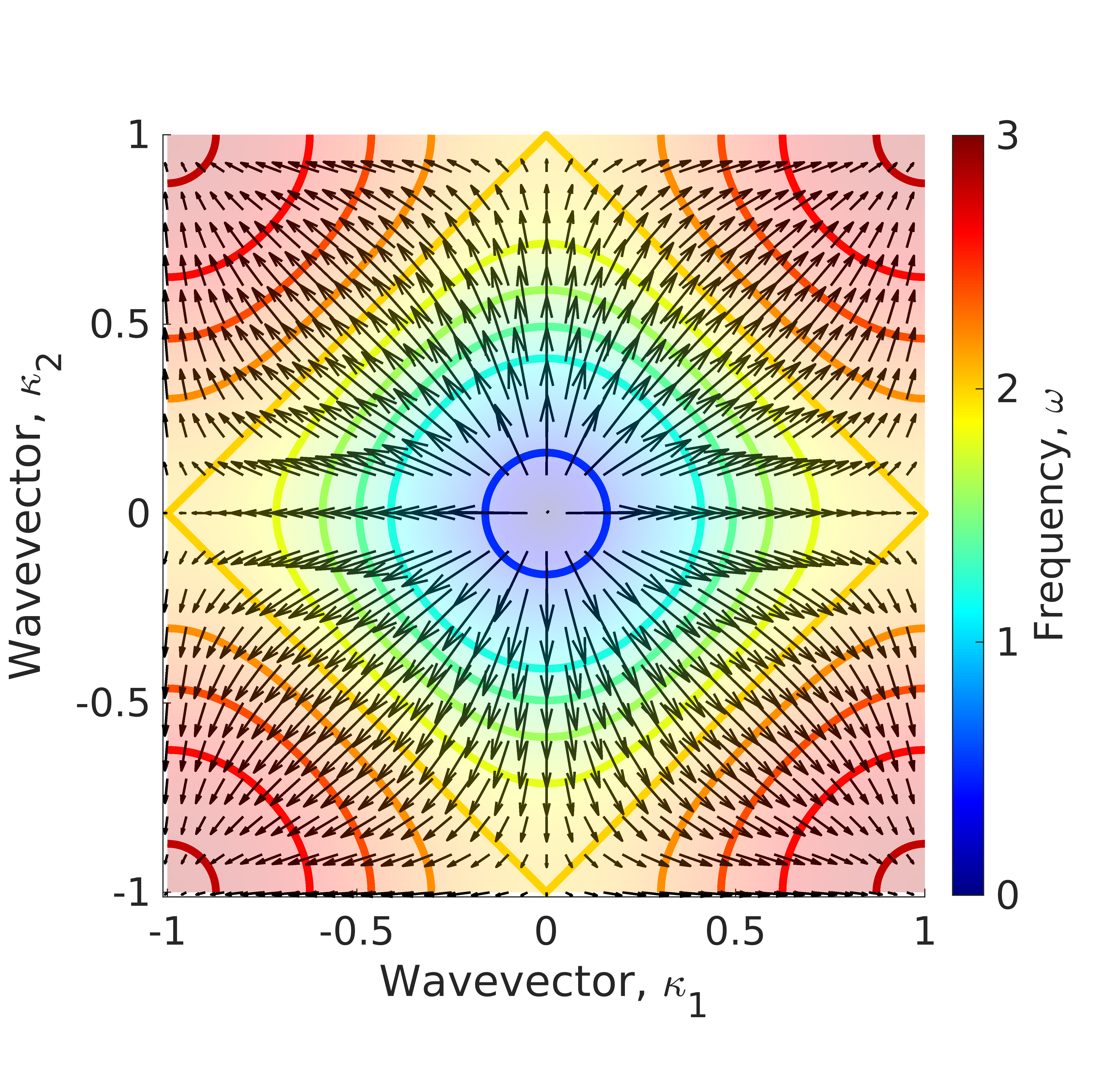}
\caption{Iso-frequency dispersion contours of the 2-D spring-mass system with local (nearest-neighbor) interactions only, $N=M=1$.
The spring stiffness values are $k_{x_1}=k_{y_1}=1$, all mass values are $m=1$,  and the lattice constant has the value $a=\pi$.
The black arrows denote the group-velocity vectors as a function of the wavevectors $(\kappa_1, \kappa_2)$.
Colorbar denotes frequency ($\omega$) values.
}
\label{fig:Local_lattice_contours}
\end{figure}
For both the examples in 2-D, we start by considering a spring-mass system  with local interactions (nearest-neighbor interactions) only, i.e., $N=M=1$, and $k_{x_1}=k_{y_1}=1, m=1,$ and $a=\pi$.
The initial displacement $u_0(x,y)$ of the masses is given by a Gaussian function of the form,
\begin{equation}
    u_0(x,y) =
    e^{-(x^2+y^2)/1000}\sin{(\kappa_1 x)}\sin{(\kappa_2 y)}.
\end{equation}
For the \textit{wave-freezing} problem we choose $(\kappa_1, \kappa_2) = (0.5, 0.5)$, and for the problem of anomalous temporal refraction we choose $(\kappa_1, \kappa_2) = (0.5, 0.75)$.
The initial wave-packet splits into four packets, propagating in the four quadrants of the $X-Y$ axes. 
For brevity, using the symmetry of the problem (without any loss of generality),
the numerical results in the main manuscript for the wave propagation are shown in the positive $X$ and positive $Y$ domain only.

Figure \ref{fig:Local_lattice_contours} shows the iso-frequency dispersion contours of the 2-D spring-mass system with local (nearest-neighbor) interactions only, that is used in both these examples.
The colorbar represents frequency ($\omega$) values.
The black arrows represent the group-velocity vectors as a function of $(\kappa_1, \kappa_2)$.
The arrows show that  for $\kappa_1, \kappa_2 \geq 0$, the group-velocity is positive everywhere.

\begin{figure}[tp!]
        \centering
        \includegraphics[width=\linewidth]{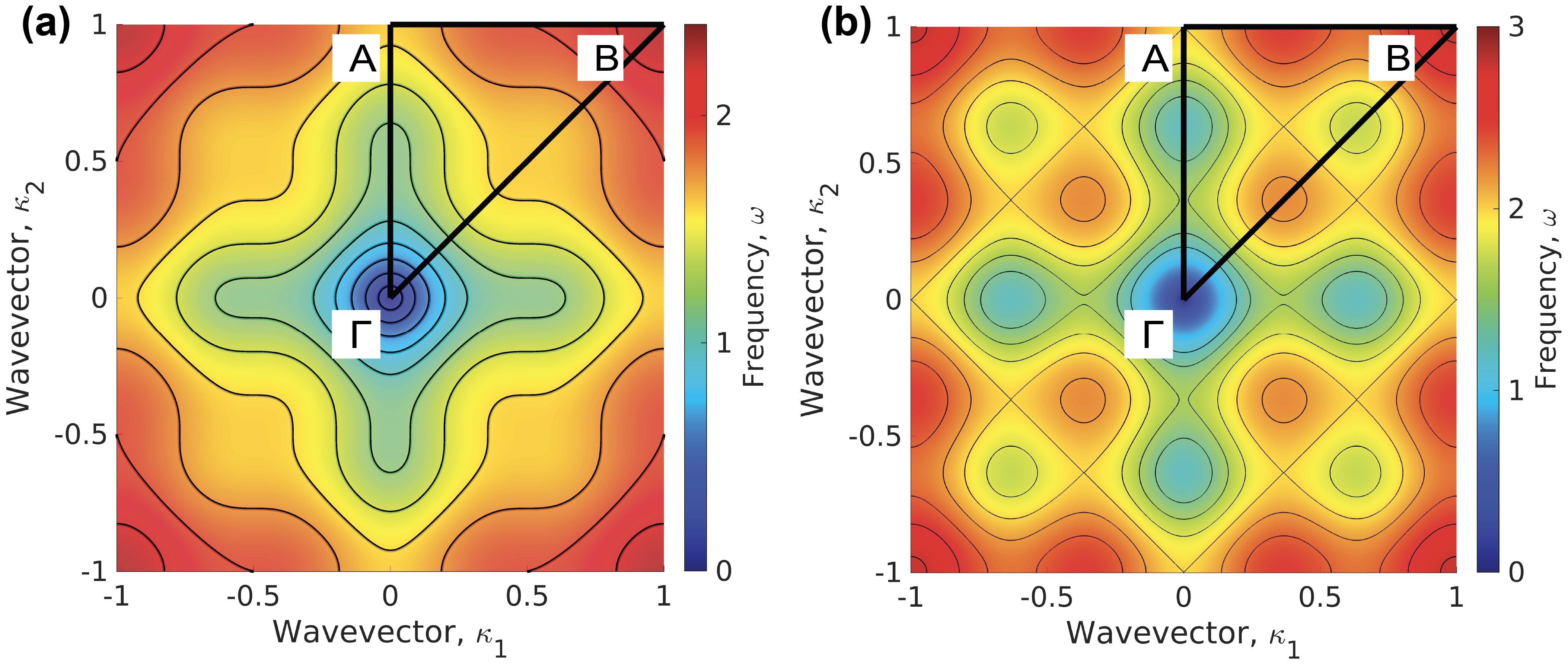}
\caption{Dispersion surface colormaps: 
(a) Colormap of the dispersion surface of the nonlocal spring-mass system used in the 2-D wave-freezing example in Fig. \red{3} of the main manuscript. 
The first BZ boundary is shown by the solid black triangle $\Gamma-A-B$.
Thin black curves represent the iso-frequency contours.
(b) Colormap of the dispersion surface of the nonlocal spring-mass system used in the 2-D anomalous temporal refraction example in Fig. \red{4} of the main manuscript. 
The first BZ boundary is shown by the solid black triangle $\Gamma-A-B$.
}
\label{fig:2d_Dispersion_surface_paths}
\end{figure}
Figure \ref{fig:2d_Dispersion_surface_paths}\red{(a)} and Fig. \ref{fig:2d_Dispersion_surface_paths}\red{(b)} show the iso-frequency dispersion contours of the nonlocal spring-mass systems used for the 2-D examples of \textit{wave-freezing} and anomalous temporal refraction. 
The irreducible Brillouin Zone (BZ) is represented by the black triangle.
In Figs. \red{3(a)} and \red{4(a)} of the main manuscript, the frequency ($\omega$) values are plotted  for wavevector values along the boundary of these Brillouin Zones shown here.
The points of the  triangle (BZ) are $\Gamma=(0,0)$, $B=(1,1)$, and $A=(0,1)$.

%%%%%%%%%%%%%%%%%%%%
%%%%%%%%%%%%%%%%%%%%
%%%%%%%%%%%%%%%%%%%%
%%%%%%%%%%%%%%%%%%%%
%%%%%%%%%%%%%%%%%%%%
%%%%%%%%%%%%%%%%%%%%
\section{Practical realization of temporal metasurfaces}
The time-modulation of material properties of the temporal metasurfaces discussed in this work may be easily achieved in an analogous electrical setting by constructing a transmission line model using inductors, capacitors, and switches, and assuming negligible loss due to the resistance.
In an equivalent electrical circuit, the inductor represents a mass and the capacitor represents a spring. 
The inductance $L$ of an inductor is analogous to the mass value $m$, and the capacitance $C$ of the capacitor is analogous to the reciprocal of the spring stiffness value $k$.
The temporal metasurfaces considered in this work require changing the stiffness values of the local and nonlocal springs in the system, with the masses being unchanged. 
It is relatively easier to change the capacitance values in an electrical circuit than changing the spring stiffness values in a spring-mass system, and the electrical circuit gives a more precise control for time-modulation of properties. 
For more details on transmission line models, the reader is referred to the work of Lurie \textit{et al.} \cite{lurie2007introduction,Lurie2016}, where transmission line models were used to model space-time laminates and space-time checkerboards.

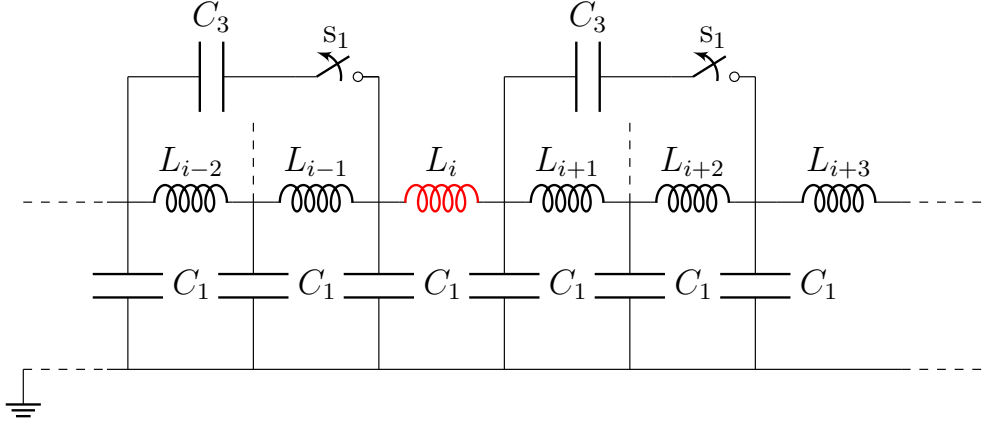
\begin{figure}[tp!]
    \centering
    \resizebox{0.8\columnwidth}{!}{%
        \begin{circuitikz} 
            \draw[dashed] (0,0) -- (1,0);
            \draw (1.25,0) to[C, l=$C_1$] (1.25,-2);
            \draw (1,-2) -- (10.5,-2);
            \draw[dashed] (10.5,-2) -- (11.5,-2);
            \draw[dashed] (0,-2) --(1,-2);
            \draw (0,-2) node[ground] {};
            
            \draw (1,0) -- (1.5,0);
            \draw (1.5,0) to[L, l=$L_{i-2}$] (2.5,0);
            \draw (2.75,0) to[C, l=$C_1$] (2.75,-2);
            % \draw (2.75,-2) to[C, l=$C_3$] (2.75,-3.5);
            % \draw (2.75,-3.5) node[ground] {};
            
            \draw (2.5,0) -- (3,0);
            \draw[dashed] (2.75,0) -- (2.75,1);
            \draw (3,0) to[L, l=$L_{i-1}$] (4,0);
            \draw (4.25,0) to[C, l=$C_1$] (4.25,-2);
            \draw (4,0) -- (4.5,0);
            \draw (4.25,0) -- (4.25,1.5);
            \draw (3.25,1.5) to[ospst=${\rm s_1}$] (4.25,1.5);
            \draw (4.25, 1.5) to[short,-o] (4, 1.5);
            \draw (1.25,1.5) to[C, l=$C_3$]  (3.25, 1.5);
            \draw (1.25,1.5) -- (1.25,0);
           
            % the i^th mass
            \draw (4.5,0) to[L, l=$L_i$, color=red] (5.5,0);
            \draw (5.75,0) to[C, l=$C_1$] (5.75,-2);
            \draw (5.75,0) to (5.75,1.5);
            \draw (5.75, 1.5) to[C, l=$C_3$] (7.75,1.5);
            \draw (7.75, 1.5) to[ospst=${\rm s_1}$] (8.75,1.5);
            \draw (8.5, 1.5) to[short,-o] (8.5, 1.5);
            \draw (8.75,1.5) to (8.75,0);

            % \draw (7,-2) to[C, l=$C_3$] (7,-3.5);
            % \draw (7,-3.5) node[ground] {};
            \draw (5.5,0) -- (6,0);
            \draw (6,0) to[L, l=$L_{i+1}$] (7,0);
            \draw[dashed] (7.25,0) -- (7.25,1);
            \draw (7.25,0) to[C, l=$C_1$] (7.25,-2);

            \draw (7,0) -- (7.5,0);
            \draw (7.5,0) to[L, l=$L_{i+2}$] (8.5,0);
            \draw (8.75,0) to[C, l=$C_1$] (8.75,-2);
            
            \draw (8.5,0) -- (9,0);
            \draw (9,0) to[L, l=$L_{i+3}$] (10.5,0);
            \draw[dashed] (10.5,0) -- (11.5,0);
        \end{circuitikz}    
        }
    \caption{A proposed transmission line model analogous to the 1-D spring-mass system used for wave-freezing in Fig. \red{1(a)} of the main manuscript.
    The $i^{th}$ mass $m_i$ is identical to the inductor $L_i$, and the spring is analogous to a capacitor (with spring stiffness value analogous to the reciprocal of the capacitance).
    Here, the local and nonlocal interactions of the mass $m_i$ with its nearest and third-nearest-neighbors are highlighted by showing equivalent connections in the LC circuit for the inductor $L_i$ (shown in red).
    Using the switches $s_1$, the system can transition from a system with only nearest-neighbor interactions to a system with nearest and third-nearest neighbor interactions, and vice-versa.
    }
    \label{fig:LC circuit}
\end{figure}

Here, we propose an example transmission line model, shown in Fig. \ref{fig:LC circuit}, that is analogous to the spring-mass system in Fig. \textcolor{red}{1(a)} of the main manuscript.
To mimic this spring-mass system, we need to be able to introduce third-nearest neighbor interactions in the transmission line.
This can be obtained by connecting the $i^{th}$ inductor with inductance $L_i$ to the inductors $L_{i+3}$ and $L_{i-3}$ by a capacitor with capacitance $C_3$.
By closing the switches $s_1$, we get a system analogous to the spring-mass system with nearest and third-nearest neighbor interactions, and by opening the switches $s_1$ we get an analogous system to the spring-mass system with nearest-neighbor interactions only.

Finally, it is worth noting that G. W. Milton and O. Mattei\cite{milton2017field,mattei2018field} introduced \textit{field patterns}, which is a new type of wave possible in space-time microstructures that does not result in chaotic disturbances.
With regards to the wave-freezing problem considered in this work, for future work it is interesting to investigate if one can find frozen field patterns and pose inverse design problems to find the associated space-time microstructures.

\bibliography{SuppInfoDoc}% Produces the bibliography via BibTeX.